\documentclass[epj]{svjour}
\usepackage{graphicx}
\usepackage{float}% floating figures at the exact position
\usepackage{amsmath}
\usepackage{amssymb}
\usepackage{verbatim}%for multiple line commenting
\usepackage{array}
\newcolumntype{P}[1]{>{\centering\arraybackslash}p{#1}}
\usepackage[numbers]{natbib}
\usepackage{tikz}
\usetikzlibrary{calc,matrix}
\usepackage{hyperref}
\usepackage{xcolor}
\hypersetup{
    colorlinks,
    linkcolor={red!50!black},
    citecolor={blue!50!black},
    urlcolor={blue!80!black}
}
\usepackage[T1]{fontenc}
\usepackage{textcomp}
\usepackage{listings}
\newcommand{\txm} {\text{T}\chi\text{M}}
\newcommand{\tpm} {\text{T}\phi\text{M}}
\newcommand{\om} {\omega}
\newcommand{\ob} {\bar{\omega}}
\newcommand{\tb} {\bar{3}}
\newcommand{\ab} {\bar{3}^{(0,1)}}
\newcommand{\bb} {\bar{3}^{(1,0)}}
\newcommand{\cb} {\bar{3}^{(1,1)}}
\newcommand{\xb} {\bar{6}}
\newcommand{\A} {\bar{a}}
\newcommand{\B} {\bar{b}}

\begin{document}
\title{Fully Constrained Majorana Neutrino Mass Matrices Using $\boldsymbol{\Sigma(72\times3)}$}
\author{R.~Krishnan\inst{1}\thanks{\href{https://orcid.org/0000-0002-0707-3267}{orcid.org/0000-0002-0707-3267}}, P.~F.~Harrison\inst{1}\thanks{\href{mailto:p.f.harrison@warwick.ac.uk} {p.f.harrison@warwick.ac.uk}}, W.~G.~Scott\inst{3}\thanks{\href{mailto:w.g.scott@rl.ac.uk}{w.g.scott@rl.ac.uk}}% etc
}                     % Do not remove
\institute{University of Warwick, Coventry, CV4 7AL, United Kingdom\and Rutherford Appleton Laboratory, Chilton, Didcot OX11 0QX, United Kingdom}  
\date{Received: date / Revised version: date}
%\and The School of the Good Shepherd, Thiruvananthapuram, 695017, India
\abstract{
In 2002, two neutrino mixing ansatze having trimaximally-mixed middle ($\nu_2$) columns, namely tri-chi-maximal mixing ($\txm$) and tri-phi-maximal mixing ($\tpm$), were proposed. In 2012, it was shown that $\txm$ with $\chi=\pm \frac{\pi}{16}$ as well as $\tpm$ with $\phi = \pm \frac{\pi}{16}$ leads to the solution, $\sin^2 \theta_{13} = \frac{2}{3} \sin^2 \frac{\pi}{16}$, consistent with the latest measurements of the reactor mixing angle, $\theta_{13}$. To obtain $\txm_{(\chi=\pm \frac{\pi}{16})}$ and $\tpm_{(\phi=\pm \frac{\pi}{16})}$, the type~I see-saw framework with fully constrained Majorana neutrino mass matrices was utilised. These mass matrices also resulted in the neutrino mass ratios, $m_1:m_2:m_3=\frac{\left(2+\sqrt{2}\right)}{1+\sqrt{2(2+\sqrt{2})}}:1:\frac{\left(2+\sqrt{2}\right)}{-1+\sqrt{2(2+\sqrt{2})}}$. In this paper we construct a flavour model based on the discrete group $\Sigma(72\times3)$ and obtain the aforementioned results. A Majorana neutrino mass matrix (a symmetric $3\times3$ matrix with 6 complex degrees of freedom) is conveniently mapped into a flavon field transforming as the complex 6 dimensional representation of $\Sigma(72\times3)$. Specific vacuum alignments of the flavons are used to arrive at the desired mass matrices.
\PACS{
      {14.60.Pq}{Neutrino mass and mixing}   \and
      {11.30.Hv}{Flavor symmetries}
     } % end of PACS codes
} %end of abstract
\maketitle
%----------------------------------------------------------------------------------------------------------------------------------
\section{Introduction}
\label{intro}
The neutrino mixing information is encapsulated in the unitary PMNS mixing matrix which, in the standard PDG parameterisation~\cite{Patrignani16}, is given by
\begin{equation}\label{eq:pmns}
\begin{split}
&U_\text{PMNS}=\\ &\left(\begin{matrix}
c_{12} c_{13} &s_{12} c_{13} &s_{13}e^{-i\delta} \\
-s_{12} c_{23}-c_{12} s_{23} s_{13} e^{i\delta} &c_{12}c_{23} -s_{12}s_{23} s_{13} e^{i\delta} &s_{23} c_{13} \\
s_{12} s_{23}-c_{12} c_{23} s_{13}e^{i\delta} &-c_{12}s_{23}-s_{12}c_{23}s_{13}e^{i\delta} &c_{23}c_{13}
\end{matrix}\right)\\ &\,\,\,\,\,\,\,\,\,\,\,  \times \text{diag}(1,\, e^{i\frac{\alpha_{21}}{2}},\, e^{i\frac{\alpha_{31}}{2}})
\end{split}
\end{equation}
where $s_{ij}=\sin \theta_{ij}, c_{ij}=\cos \theta_{ij}$. The three mixing angles $\theta_{12}$ (solar angle), $\theta_{23}$ (atmospheric angle) and $\theta_{13}$ (reactor angle) along with the $CP$-violating complex phases (the Dirac phase, $\delta$, and the two Majorana phases, $\alpha_{21}$ and $\alpha_{31}$) parameterise $U_{PMNS}$. In comparison to the small mixing angles observed in the quark sector, the neutrino mixing angles are found to be relatively large~\cite{Esteban17}:
\begin{align}
\sin^2 \theta_{12} &= 0.271\rightarrow 0.345 \,,\label{eq:anglevalues1}\\
\sin^2 \theta_{23} &= 0.385\rightarrow 0.635 \,,\label{eq:anglevalues2}\\
\sin^2 \theta_{13} &= 0.01934\rightarrow 0.02392\,. \label{eq:anglevalues3}
\end{align}
The values of the complex phases are unknown at present. Besides measuring the mixing angles, the neutrino oscillation experiments also proved that neutrinos are massive particles. These experiments measure the mass-squared differences of the neutrinos and currently their values are known to be ~\cite{Esteban17},
\begin{gather}
\Delta m_{21}^2=70.3\rightarrow 80.9~\text{meV}^2,\label{eq:massvalues1}\\
|\Delta m_{31}^2|=2407 \rightarrow 2643~\text{meV}^2.\label{eq:massvalues2}
\end{gather}

Several mixing ansatze with a trimaximally-mixed second column for $U_\text{PMNS}$, i.e.~$|U_{e2}|=|U_{\mu2}|=|U_{\tau2}|=\frac{1}{\sqrt{3}}$, were proposed during the early 2000s~\cite{Harrison99, Harrison02b, Harrison02, Xing02, Harrison03}. Here we briefly revisit two of those, the tri-chi-maximal mixing ($\txm$) and the tri-phi-maximal mixing ($\tpm$), which are relevant to our model. They can be conveniently parameterised~\cite{Harrison02} as follows   
\begin{align}
U_{\txm}&=\left(\begin{matrix}\sqrt{\frac{2}{3}}\cos \chi & \frac{1}{\sqrt{3}} & \sqrt{\frac{2}{3}}\sin \chi\\
-\frac{\cos \chi}{\sqrt{6}}-i\frac{\sin \chi}{\sqrt{2}} & \frac{1}{\sqrt{3}} & i\frac{\cos \chi}{\sqrt{2}}-\frac{\sin \chi}{\sqrt{6}}\\
-\frac{\cos \chi}{\sqrt{6}}+i\frac{\sin \chi}{\sqrt{2}} & \frac{1}{\sqrt{3}} & -i\frac{\cos \chi}{\sqrt{2}}-\frac{\sin \chi}{\sqrt{6}}
\end{matrix}\right),\label{eq:txmform}\\
U_{\tpm}&=\left(\begin{matrix}\sqrt{\frac{2}{3}}\cos \phi & \frac{1}{\sqrt{3}} & \sqrt{\frac{2}{3}}\sin \phi\\
-\frac{\cos \phi}{\sqrt{6}}-\frac{\sin \phi}{\sqrt{2}} & \frac{1}{\sqrt{3}} & \frac{\cos \phi}{\sqrt{2}}-\frac{\sin \phi}{\sqrt{6}}\\
-\frac{\cos \phi}{\sqrt{6}}+\frac{\sin \phi}{\sqrt{2}} & \frac{1}{\sqrt{3}} & -\frac{\cos \phi}{\sqrt{2}}-\frac{\sin \phi}{\sqrt{6}}
\end{matrix}\right)\label{eq:tpmform}.
\end{align}
Both $\txm$ and $\tpm$\footnote{$TM_i$ ($TM^i$) has been proposed~\cite{TM2a, TM2b} as a nomenclature to denote the mixing matrices that preserve various columns (rows) of the tribimaximal mixing~\cite{Harrison02b}. Under this notation, both $\txm$ and $\tpm$ fall under the category of $TM_2$. To be more specific, $TM_2$ which breaks $CP$ maximally is $\txm$ and $TM_2$ which conserves $CP$ is $\tpm$.}  have one free parameter each ($\chi$ and $\phi$) which directly corresponds to the reactor mixing angle, $\theta_{13}$, through the $U_{e3}$ elements of the mixing matrices. The three mixing angles and the Dirac $CP$ phase obtained by relating Eq.~(\ref{eq:pmns}) with Eqs.~(\ref{eq:txmform},~\ref{eq:tpmform}) are shown in Table~\ref{tab:anglesandphase}.
{\renewcommand{\arraystretch}{1.6}
\begin{table}[]
\begin{center}
\begin{tabular}{|c|c|c|c|c|}
%\hline
\hline
	&$\sin^2 \theta_{13}$	&$\sin^2 \theta_{12}$	&$\sin^2 \theta_{23}$	&$\delta$	\\
%\hline
\hline
$\txm$	&$\frac{2}{3} \sin^2 \chi$	&$\frac{1}{\left(3-2\sin^2 \chi\right)}$	&$\frac{1}{2}$	&$\pm\frac{\pi}{2}$	\\[5pt]
%\hline
$\tpm$	&$\frac{2}{3} \sin^2 \phi$	&$\frac{1}{\left(3-2\sin^2 \phi\right)}$	&$\frac{2 \sin^2 \left(\frac{2\pi}{3}+\phi\right)}{\left(3-2\sin^2 \phi\right)}$	&$0,~\pi$	\\[5pt]
%\hline
\hline
\end{tabular}
\end{center}
\caption{The standard PDG observables $\theta_{13}$, $\theta_{12}$, $\theta_{23}$ and $\delta$ in terms of the parameters $\chi$ and $\phi$. Note that the range of $\chi$ as well as $\phi$ is $-\frac{\pi}{2}$ to $+\frac{\pi}{2}$. In $\txm$ ($\tpm$), the parameter $\chi$ ($\phi$) being in the first and the fourth quadrant correspond to $\delta$ equal to $+\frac{\pi}{2}$ ($0$) and $-\frac{\pi}{2}$ ($\pi$) respectively.}
\label{tab:anglesandphase}
\end{table}}
\noindent In $\txm$, since $\delta = \pm \frac{\pi}{2}$, $CP$ violation is maximal for a given set of mixing angles. The Jarlskog $CP$-violating invariant~\cite{JCP1, JCP2, JCP3, JCP4, JCP5} in the context of $\txm$~\cite{Harrison02} is given by 
\begin{equation}\label{eq:jcp}
J=\frac{\sin 2\chi}{6\sqrt{3}}.
\end{equation} 
On the other hand, $\tpm$ is $CP$ conserving, i.e.~$\delta = 0,~\pi$, and thus $J=0$. 

Since the reactor angle was discovered to be non-zero at the Daya Bay reactor experiment in 2012~\cite{DayaBay}, there has been a resurgence of interest~\cite{Feruglio13, TM22, TM23, TM24, Feruglio14, S4Paper, LIS, SteveFour, Delta16, Thomas, Vien16, Vien16b} in $\txm$ and $\tpm$ and their equivalent forms. For any $CP$-conserving ($\delta = 0,~\pi$) mixing matrix with non-zero $\theta_{13}$ and trimaximally-mixed $\nu_2$ column, we can have an equivalent parameterisation realised using the $\tpm$ matrix. Here the ``equivalence'' is with respect to the neutrino oscillation experiments. The oscillation scenario is completely determined by the three mixing angles and the Dirac phase (Majorana phases are not observable in neutrino oscillations), i.e.~we have a total of four degrees of freedom in the mixing matrix. If we assume $CP$ conservation and also assume that the $\nu_2$ column is trimaximally mixed, then there is only one degree of freedom left. It is exactly this degree of freedom which is parameterised using $\phi$ in $\tpm$ mixing. Similarly any mixing matrix with $\delta = \pm \frac{\pi}{2}$, $\theta_{13}\neq0$ and trimaximal $\nu_2$ column is equivalent to $\txm$ mixing.

In 2012~\cite{LIS}, shortly after the discovery of the non-zero reactor mixing angle, it was shown that $\txm_{(\chi=\pm \frac{\pi}{16})}$ as well as $\tpm_{(\phi=\pm \frac{\pi}{16})}$ results in a reactor mixing angle, 
\begin{equation}\label{eq:theta13}
\begin{split}
\sin^2 \theta_{13} &= \frac{2}{3} \sin^2 \frac{\pi}{16} \\
&= 0.025,
\end{split}
\end{equation} 
consistent with the experimental data. The model was constructed in the Type-1 see-saw framework~\cite{seesaw2,seesaw3,seesaw4,seesaw5}. Four cases of Majorana mass matrices were discussed:
\begin{align}
M_\text{Maj} & \propto \left(\begin{matrix}(2-\sqrt{2}) & 0 & \frac{1}{\sqrt{2}}\\
0 & 1 & 0\\
\frac{1}{\sqrt{2}} & 0 & 0
\end{matrix}\right),& &\propto\left(\begin{matrix}0 & 0 & \frac{1}{\sqrt{2}}\\
0 & 1 & 0\\
\frac{1}{\sqrt{2}} & 0 & (2-\sqrt{2})
\end{matrix}\right),\label{eq:txmmat}\\
M_\text{Maj} & \propto \left(\begin{matrix}i+\frac{1-i}{\sqrt{2}} & 0 & 1-\frac{1}{\sqrt{2}}\\
0 & 1 & 0\\
1-\frac{1}{\sqrt{2}} & 0 & -i+\frac{1+i}{\sqrt{2}}
\end{matrix}\right),& &\propto\left(\begin{matrix}-i+\frac{1+i}{\sqrt{2}} & 0 & 1-\frac{1}{\sqrt{2}}\\
0 & 1 & 0\\
1-\frac{1}{\sqrt{2}} & 0 & i+\frac{1-i}{\sqrt{2}}
\end{matrix}\right)\label{eq:tpmmat}
\end{align}
where $M_\text{Maj}$ is the coupling among the right-handed neutrino fields, i.e.~$\overline{(\nu_R)^c}M_\text{Maj} \nu_R$. In Ref.~\cite{LIS}, the mixing matrix was modelled in the form
\begin{equation}
U_\text{PMNS} = \frac{1}{\sqrt{3}}\left(\begin{matrix}1 & 1 & 1\\
1 & \omega & \bar{\omega}\\
1 & \bar{\omega} & \omega
\end{matrix}\right)U_\nu\,, \quad \text{with } \omega=e^{i\frac{2\pi}{3}} \text{ and } \bar{\omega}=e^{\text{-}i\frac{2\pi}{3}},
\end{equation}
in which the $3\times3$ trimaximal contribution came from the charged-lepton sector. $U_\nu$, on the other hand, was the contribution from the neutrino sector. The four $U_\nu$s vis-a-vis the four Majorana neutrino mass matrices given in Eqs.~(\ref{eq:txmmat}) and Eqs.~(\ref{eq:tpmmat}), gave rise to $\txm_{(\chi=\pm \frac{\pi}{16})}$ and $\tpm_{(\phi=\pm \frac{\pi}{16})}$ respectively. ~All the four mass matrices, Eqs.~(\ref{eq:txmmat},~\ref{eq:tpmmat}), have the eigenvalues $\frac{1+\sqrt{2(2+\sqrt{2})}}{\left(2+\sqrt{2}\right)}$, $1$ and $\frac{-1+\sqrt{2(2+\sqrt{2})}}{\left(2+\sqrt{2}\right)}$. Due to the see-saw mechanism, the neutrino masses become inversely proportional to the eigenvalues of the Majorana mass matrices, resulting in the neutrino mass ratios
\begin{equation}\label{eq:numass}
m_1:m_2:m_3=\frac{\left(2+\sqrt{2}\right)}{1+\sqrt{2(2+\sqrt{2})}}:1:\frac{\left(2+\sqrt{2}\right)}{-1+\sqrt{2(2+\sqrt{2})}}\,\,.
\end{equation}
Using these ratios and the experimentally-measured mass-squared differences, the light neutrino mass was predicted to be around $25~\text{meV}$.

In this paper we use the discrete group $\Sigma(72\times3)$ to construct a flavon model that essentially reproduces the above results. Unlike in Ref.~\cite{LIS} where the neutrino mass matrix was decomposed into a symmetric product of two matrices, here a single sextet representation of the flavour group is used to build the neutrino mass matrix. A brief discussion of the group $\Sigma(72\times3)$ and its representations is provided in Section~2 of this paper. Appendix~A contains more details such as the tensor product expansions of its various irreducible representations (irreps) and the corresponding Clebsch-Gordan (C-G) coefficients. In Section~3, we describe the model with its fermion and flavon field content in relation to these irreps. Besides the aforementioned sextet flavon, we also introduce triplet flavons in the model to build the charged-lepton mass matrix. The flavons are assigned specific vacuum expectation values (VEVs) to obtain the required neutrino and charged-lepton mass matrices. A detailed description of how the charged-lepton mass matrix attains its hierarchical structure is deferred to Appendix~B. In Section~4, we obtain the phenomenological predictions and compare them with the current experimental data along with the possibility of further validation from future experiments. Finally the results are summarised in Section~5. The construction of suitable flavon potentials which generate the set of VEVs used in our model is demonstrated in Appendix~C. 
%----------------------------------------------------------------------------------------------------------------------------------
\section{The Group $\Sigma(72\times3)$ and its Representations}

Discrete groups have been used extensively in the description of flavour symmetries. Historically, the study of discrete groups can be traced back to the study of symmetries of geometrical objects. Tetrahedron, cube, octahedron, dodecahedron and icosahedron, which are the famous Platonic solids, were known to the ancient Greeks. These objects are the only regular polyhedra with congruent regular polygonal faces. Interestingly, the symmetry groups of the platonic solids are the most studied in the context of flavour symmetries too - $A_4$ (tetrahedron), $S_4$ (cube and its dual octahedron) and $A_5$ (dodecahedron and its dual icosahedron). These polyhedra live in the three-dimensional Euclidean space. In the context of flavour physics, it might be rewarding to study similar polyhedra that live in three-dimensional complex Hilbert space. In fact, five such complex polyhedra that correspond to the five Platonic solids exist as shown by Coxeter~\cite{Coxpoly}. They are $3\{3\}3\{3\}3$, $2\{3\}2\{4\}p$, $p\{4\}2\{3\}2$, $2\{4\}3\{3\}3$, $3\{3\}3\{4\}2$ where we have used the generalised schlafli symbols~\cite{Coxpoly} to represent the polyhedra. The polyhedron $3\{3\}3\{3\}3$ known as the Hessian polydehron can be thought of as the tetrahedron in the complex space. Its full symmetry group has 648 elements and is called $\Sigma(216\times3)$. Like the other discrete groups relevant in flavour symmetry, $\Sigma(216\times3)$ is also a subgroup of the continuous group $U(3)$. 

The principal series of $\Sigma(216\times3)$~\cite{Sigma1} is given by
\begin{equation}
\{e\} \triangleleft Z_3 \triangleleft \Delta(27) \triangleleft \Delta(54) \triangleleft \Sigma(72\times3) \triangleleft \Sigma(216\times3).
\end{equation}
Our flavour symmetry group, $\Sigma(72\times3)$, is the maximal normal subgroup of $\Sigma(216\times3)$. So we get $\Sigma(216\times3)/\Sigma(72\times3)=Z_3$. Various details about the properties of the group $\Sigma(72\times3)$ and its representations can be found in Refs.~\cite{Sigma1, Sigma2, Smallgroup, SigmaHagedorn, Merle}. Note that $\Sigma(72\times3)$ is quite distinct from $\Sigma(216)$ which is defined using the relation $\Sigma(216\times3)/Z_3=\Sigma(216)$. In other words, $\Sigma(216\times3)$ forms the triple cover of $\Sigma(216)$. $\Sigma(216\times3)$ as well as $\Sigma(216)$ is sometimes referred to as the Hessian group. In terms of the GAP~\cite{GAP4,GAPSmallGroups} nomenclature, we have $\Sigma(216\times3)\equiv \text{SmallGroup(648,532)},\,$ $\Sigma(72\times3)\equiv \text{SmallGroup(216,88)}\,$ and $\,\Sigma(216)\equiv \text{SmallGroup(216,153)}$.

We find that, in the context of flavour physics and model building, $\Sigma(72\times3)$ has an appealing feature: it is the smallest group containing a complex three-dimensional representation whose tensor product with itself results in a complex six-dimensional representation~\footnote{We studied the comprehensive list of finite subgroups of $U(3)$ provided in Ref.~\cite{Allsubgroups} and determined that $\Sigma(72\times3)$ is the smallest group having this feature.}, i.e.
\begin{equation}\label{eq:tensor0}
\boldsymbol{3}\otimes\boldsymbol{3}=\boldsymbol{6}\oplus\boldsymbol{\tb}.
\end{equation}
With a suitably chosen basis for $\boldsymbol{6}$ we get
\begin{equation}\label{eq:tensor1exp}
\boldsymbol{6}\equiv\left(\begin{matrix}a_1 b_1 \\
	a_2 b_2\\
	a_3 b_3\\
	\frac{1}{\sqrt{2}}\left(a_2 b_3 + a_3 b_2\right)\\
	\frac{1}{\sqrt{2}}\left(a_3 b_1 + a_1 b_3\right)\\
\frac{1}{\sqrt{2}}\left(a_1 b_2 + a_2 b_1\right)
\end{matrix}\right), \quad \boldsymbol{\bar{3}} \equiv
\left(\begin{matrix}\frac{1}{\sqrt{2}}\left(a_2 b_3 - a_3 b_2\right)\\
	\frac{1}{\sqrt{2}}\left(a_3 b_1 - a_1 b_3\right)\\
\frac{1}{\sqrt{2}}\left(a_1 b_2 - a_2 b_1\right)
\end{matrix}\right)
\end{equation}
where $(a_1, a_2, a_3)^T$ and $(b_1, b_2, b_3)^T$ represent the triplets appearing in the LHS of Eq.~(\ref{eq:tensor0}). All the symmetric components of the tensor product together form the representation $\boldsymbol{6}$ and the antisymmetric components form $\boldsymbol{\tb}$. For the $SU(3)$ group it is well known that the tensor product of two $\boldsymbol{3}$s gives rise to a symmetric $\boldsymbol{6}$ and an antisymmetric $\boldsymbol{\tb}$. $\Sigma(72\times3)$ being a subgroup of $SU(3)$, of course, has its $\boldsymbol{6}$ and $\boldsymbol{\tb}$ embedded in the $\boldsymbol{6}$ and $\boldsymbol{\tb}$ of $SU(3)$.  

Consider the complex conjugation of Eq.~(\ref{eq:tensor0}), i.e.~$\boldsymbol{\tb}\otimes\boldsymbol{\tb}=\boldsymbol{\xb}\oplus\boldsymbol{3}$. Let the right-handed neutrinos form a triplet, $\nu_R=(\nu_{R1},\nu_{R2},\nu_{R3})^T$, which transforms as a $\boldsymbol{\tb}$. Symmetric (and also Lorentz invariant) combination of two such triplets leads to a conjugate sextet, $\bar{S}_\nu$, which transforms as a $\boldsymbol{\xb}$,
\begin{equation}\label{eq:Xnu}
\bar{S}_\nu= \left(\begin{matrix}\nu_{R1}.\nu_{R1}\\
	\nu_{R2}.\nu_{R2}\\
	\nu_{R3}.\nu_{R3}\\
	\frac{1}{\sqrt{2}}\left(\nu_{R2}.\nu_{R3} + \nu_{R3}.\nu_{R2}\right)\\
	\frac{1}{\sqrt{2}}\left(\nu_{R3}.\nu_{R1} + \nu_{R1}.\nu_{R3}\right)\\
\frac{1}{\sqrt{2}}\left(\nu_{R1}.\nu_{R2} + \nu_{R2}.\nu_{R1}\right)
\end{matrix}\right)\equiv \boldsymbol{\xb}
\end{equation}
where $\nu_{Ri}.\nu_{Rj}$ is the Lorentz invariant product of the right-handed neutrino Weyl spinors. We may couple $\bar{S}_\nu$ to a flavon field
\begin{equation}\label{eq:xinu}
\xi=(\xi_1,\xi_2,\xi_3,\xi_4,\xi_5,\xi_6)^T 
\end{equation}
which transforms as a $\boldsymbol{6}$ to construct the invariant term
\begin{equation}\label{eq:Tnu}
\bar{S}_\nu^T \xi = \left(\begin{matrix}\nu_{R1}\\
	\nu_{R2}\\
\nu_{R3}
\end{matrix}\right)^T. \left(\begin{matrix}\xi_1 & \frac{1}{\sqrt{2}}\xi_6 & \frac{1}{\sqrt{2}}\xi_5\\
       \frac{1}{\sqrt{2}}\xi_6 &\xi_2 & \frac{1}{\sqrt{2}}\xi_4\\
       \frac{1}{\sqrt{2}}\xi_5 & \frac{1}{\sqrt{2}}\xi_4 &\xi_3
\end{matrix}\right).\left(\begin{matrix}\nu_{R1}\\
	\nu_{R2}\\
\nu_{R3}
\end{matrix}\right).
\end{equation}
In general, the $3\times3$ Majorana mass matrix is symmetric and has six complex degrees of freedom. Therefore, using Eq.~(\ref{eq:Tnu}), any required mass matrix can be obtained through a suitably chosen vacuum expectation value (VEV) for the flavon field. 

%Constructing the symmetric Majorana neutrino mass matrix with the help of flavon sextets has been attempted before, eg. scalar fields transforming as the antisextets of $SU(3)_L$ are used in Refs.~\cite{Long1, Long2}.

{\renewcommand{\arraystretch}{1.5}
\begin{table*}[t]
\begin{center}
\scriptsize
\begin{tabular}{||c||c c c|c|c c c|c c c|c c c|c c c||}
\hline
\hline
$\Sigma(72\times3)$&$C_1$&$C_2$&$C_3$&$C_4$&$C_5$&$C_6$&$C_7$&$C_8$&$C_9$&$C_{10}$&$C_{11}$&$C_{12}$&$C_{13}$&$C_{14}$&$C_{15}$&$C_{16}$\\
$\#C_k$&$1$&$1$&$1$&$24$&$9$&$9$&$9$&$18$&$18$&$18$&$18$&$18$&$18$&$18$&$18$&$18$\\
$ord(C_k)$&$1$&$3$&$3$&$3$&$2$&$6$&$6$&$4$&$12$&$12$&$4$&$12$&$12$&$4$&$12$&$12$\\
\hline
\hline
$\boldsymbol{1}$&$1$&$1$&$1$		&$1$&$1$&$1$&$1$		&$1$&$1$&$1$		&$1$&$1$&$1$		&$1$&$1$&$1$\\

\hline
$\boldsymbol{1^{(0,1)}}$&$1$&$1$&$1$		&$1$&$1$&$1$&$1$		&$1$&$1$&$1$		&$-1$&$-1$&$-1$		&$-1$&$-1$&$-1$\\
$\boldsymbol{1^{(1,0)}}$&$1$&$1$&$1$		&$1$&$1$&$1$&$1$		&$-1$&$-1$&$-1$		&$1$&$1$&$1$		&$-1$&$-1$&$-1$\\
$\boldsymbol{1^{(1,1)}}$&$1$&$1$&$1$		&$1$&$1$&$1$&$1$		&$-1$&$-1$&$-1$		&$-1$&$-1$&$-1$		&$1$&$1$&$1$\\
\hline
$\boldsymbol{2}$&$2$&$2$&$2$		&$2$&$-2$&$-2$&$-2$		&$0$&$0$&$0$		&$0$&$0$&$0$		&$0$&$0$&$0$\\
\hline
$\boldsymbol{3}$&$3$&$3\om$&$3\ob$	&$0$&$-1$&$-\om$&$-\ob$		&$1$&$\om$&$\ob$	&$1$&$\om$&$\ob$	&$1$&$\om$&$\ob$\\
\hline
$\boldsymbol{3^{(0,1)}}$&$3$&$3\om$&$3\ob$	&$0$&$-1$&$-\om$&$-\ob$		&$1$&$\om$&$\ob$	&$-1$&$-\om$&$-\ob$	&$-1$&$-\om$&$-\ob$\\
$\boldsymbol{3^{(1,0)}}$&$3$&$3\om$&$3\ob$	&$0$&$-1$&$-\om$&$-\ob$		&$-1$&$-\om$&$-\ob$	&$1$&$\om$&$\ob$	&$-1$&$-\om$&$-\ob$\\
$\boldsymbol{3^{(1,1)}}$&$3$&$3\om$&$3\ob$	&$0$&$-1$&$-\om$&$-\ob$		&$-1$&$-\om$&$-\ob$	&$-1$&$-\om$&$-\ob$	&$1$&$\om$&$\ob$\\
\hline
$\boldsymbol{\tb}$&$3$&$3\ob$&$3\om$	&$0$&$-1$&$-\ob$&$-\om$		&$1$&$\ob$&$\om$	&$1$&$\ob$&$\om$	&$1$&$\ob$&$\om$\\
\hline
$\boldsymbol{\ab}$&$3$&$3\ob$&$3\om$&$0$&$-1$&$-\ob$&$-\om$	&$1$&$\ob$&$\om$	&$-1$&$-\ob$&$-\om$	&$-1$&$-\ob$&$-\om$\\
$\boldsymbol{\bb}$&$3$&$3\ob$&$3\om$&$0$&$-1$&$-\ob$&$-\om$	&$-1$&$-\ob$&$-\om$	&$1$&$\ob$&$\om$	&$-1$&$-\ob$&$-\om$\\
$\boldsymbol{\cb}$&$3$&$3\ob$&$3\om$&$0$&$-1$&$-\ob$&$-\om$	&$-1$&$-\ob$&$-\om$	&$-1$&$-\ob$&$-\om$	&$1$&$\ob$&$\om$\\
\hline
$\boldsymbol{6}$&$6$&$6\ob$&$6\om$&$0$&$2$&$2\ob$&$2\om$		&$0$&$0$&$0$		&$0$&$0$&$0$		&$0$&$0$&$0$\\
$\boldsymbol{\xb}$&$6$&$6\om$&$6\ob$	&$0$&$2$&$2\om$&$2\ob$		&$0$&$0$&$0$		&$0$&$0$&$0$		&$0$&$0$&$0$\\
\hline
$\boldsymbol{8}$&$8$&$8$&$8$		&$-1$&$0$&$0$&$0$		&$0$&$0$&$0$		&$0$&$0$&$0$		&$0$&$0$&$0$\\
\hline
\hline
\end{tabular}
\end{center}
\caption{Character table of $\Sigma(72\times3)$.}
\label{tab:charactertable}
\end{table*} 
}

To describe the representation theory of $\Sigma(72\times3)$ we largely follow Ref.~\cite{Sigma1}. $\Sigma(72\times3)$ can be constructed using four generators, namely $C$, $E$, $V$ and $X$~\cite{Sigma1}. For the three-dimensional representation, we have
\begin{equation}\label{eq:gen3}
\begin{split}
&C \equiv
\left(\begin{matrix}1 & 0 & 0\\
       0 & \om & 0\\
       0 & 0 & \ob
\end{matrix}\right), \quad \quad \quad \, \, \, E \equiv
\left(\begin{matrix}0 & 1 & 0\\
       0 & 0 & 1\\
       1 & 0 & 0
\end{matrix}\right),\\
&V\equiv
-\frac{i}{\sqrt{3}}\left(\begin{matrix}1 & 1 & 1\\
       1 & \om & \ob\\
       1 & \ob & \om
\end{matrix}\right), \quad X\equiv
-\frac{i}{\sqrt{3}}\left(\begin{matrix}1 & 1 & \ob\\
       1 & \om & \om\\       
       \om & 1 & \om
       \end{matrix}\right).
\end{split}
\end{equation}
The characters of the irreducible representations of $\Sigma(72\times3)$ are given in Table~\ref{tab:charactertable}. Tensor product expansions of various representations relevant to our model are given in Appendix~A. There we also provide the corresponding C-G coefficients and the generator matrices.

%----------------------------------------------------------------------------------------------------------------------------------
\section{The Model}
\addtocontents{toc}{\protect\setcounter{tocdepth}{1}}

In this paper we construct our model in the Standard Model framework with the addition of heavy right-handed neutrinos. Through the type~I see-saw mechanism, light Majorana neutrinos are produced. The fermion and flavon content of the model, together with the representations to which they belong, are given in Table~\ref{tab:flavourcontent}. In addition to $\Sigma(72\times3)$, we have introduced a flavour group $C_4 = \{1, -1, i, -i\}$ for obtaining the observed mass hierarchy for the charged leptons. The Standard Model Higgs field is assigned to the trivial (singlet) representation of the flavour groups.

{\renewcommand{\arraystretch}{1.5}
\begin{table}[]
\begin{center}
\begin{tabular}{|c|c c c c c c c c|}
\hline
Fermions&$e_R$	&$\mu_R$&$\tau_R$&$L$	&$\nu_R$&$\phi_\alpha$&$\phi_\beta$&$\xi$\\
\hline
$\Sigma(72\times3)$&$\boldsymbol{1}$&$\boldsymbol{1}$&$\boldsymbol{1}$&$\boldsymbol{\tb}$&$\boldsymbol{\tb}$&$\boldsymbol{3}$&$\boldsymbol{3}$&$\boldsymbol{6}$\\
$C_4$	&$-1$&$1$&$i$&$1$&$1$&$-i$&$i$&$1$\\
\hline
\end{tabular}
\end{center}
\caption{The flavour structure of the model. The three families of the left-handed-weak-isospin lepton doublets form the triplet $L$ and the three right-handed heavy neutrinos form the triplet $\nu_R$. The flavons $\phi_\alpha$, $\phi_\beta$ and $\xi$, are scalar fields and are gauge invariants. On the other hand, they transform non-trivially under the flavour groups.}
\label{tab:flavourcontent}
\end{table}
}

For the charged leptons, we obtain the mass term
\begin{equation}\label{eq:mclept}
\left(y_\tau L^\dagger \tau_R \frac{\bar{\phi}_\beta}{\Lambda}+y_\mu L^\dagger \mu_R \frac{\sqrt{2}\bar{A}_{\beta\alpha}}{\Lambda^2}\right)H+\,\,\mathcal{H}.\mathcal{T}.
\end{equation}
where $H$ is the Standard Model Higgs, $\Lambda$ is the cut-off scale, $y_\tau$ and $y_\mu$ are the coupling constants for the $\tau$-sector and the $\mu$-sector respectively. $\bar{A}_{\beta\alpha}$ is the conjugate triplet obtained from $\phi_\beta$ and $\phi_\alpha$, constructed in the same way as the second part of Eq.~(\ref{eq:tensor1exp}),
\begin{equation}\label{eq:Aab}
\bar{A}_{\beta\alpha} \equiv
\left(\begin{matrix}\frac{1}{\sqrt{2}}\left(\phi_{\beta 2} \phi_{\alpha 3} - \phi_{\beta 3} \phi_{\alpha 2}\right)\\
	\frac{1}{\sqrt{2}}\left(\phi_{\beta 3} \phi_{\alpha 1} - \phi_{\beta 1} \phi_{\alpha 3}\right)\\
\frac{1}{\sqrt{2}}\left(\phi_{\beta 1} \phi_{\alpha 2} - \phi_{\beta 2} \phi_{\alpha 1}\right)
\end{matrix}\right)
\end{equation}
where $\phi_\alpha=(\phi_{\alpha 1}, \phi_{\alpha 2}, \phi_{\alpha 3})^T$ and $\phi_\beta=(\phi_{\beta 1}, \phi_{\beta 2}, \phi_{\beta 3})^T$. 

$L^\dagger \tau_R$ transforms as $\boldsymbol{3}\times i$ under the flavour group, $\Sigma(72\times3)\times C_4$. The flavon $\bar{\phi}_\beta$ transforms as $\boldsymbol{\tb}\times -i$ and hence it couples to $L^\dagger \tau_R$ as shown in Eq.~(\ref{eq:mclept}). No other coupling involving $\tau_R$, $\mu_R$ or $e_R$ with either $\bar{\phi}_\beta$ or $\bar{\phi}_\alpha$ is allowed, given the $C_4$ assignments in Table~\ref{tab:flavourcontent}. However, $L^\dagger \mu_R$ and $\bar{A}_{\beta\alpha}$, which transform as $\boldsymbol{3}\times 1$ and $\boldsymbol{\tb}\times 1$ respectively, can couple, Eq.~(\ref{eq:mclept}). Note that $\bar{A}_{\beta\alpha}$ is a second order product of $\phi_\beta$ and $\phi_\alpha$ and it is antisymmetric. No other second order product transforming as $\boldsymbol{\tb}$ exists, since the antisymmetric product of $\phi_\beta$ with itself or $\phi_\alpha$ with itself vanishes. $\mathcal{H}.\mathcal{T}.$ represents all the higher order terms, i.e.~ the terms consisting of higher order products of the flavons, coupling to $e_R$, $\mu_R$ and $\tau_R$. It can be shown that, for obtaining a flavon term coupling to the $e_R$, we require at least quartic order\footnote{Refer to Appendix~B for an analysis of the higher order products of $\phi_\alpha$ and $\phi_\beta$.}.

The VEV of the Higgs, $(0, h_o)$, breaks the weak gauge symmetry. For the flavons $\bar{\phi}_\alpha$ and $\bar{\phi}_\beta$, we assign the vacuum alignments\footnote{Refer to Appendix~C for the details of the flavon potential that leads to these VEVs.} \addtocounter{footnote}{-2}\addtocounter{Hfootnote}{-2}
\begin{equation}\label{eq:leptvev}
\langle\bar{\phi}_\alpha\rangle=V^\dagger(1,0,0)^Tm,\quad\langle\bar{\phi}_\beta\rangle=V^\dagger(0,0,1)^Tm
\end{equation}
where $V$ is one of the generators of $\Sigma(72\times3)$ given in Eqs.~(\ref{eq:gen3}) and is proportional to the $3\times3$ trimaximal matrix. The constant $m$ has dimensions of mass. Substituting these vacuum alignments in Eq.~(\ref{eq:mclept}) leads to the following charged-lepton mass term
\begin{equation}\label{eq:leptcontrib}
\left(\begin{matrix} e_L\\
\mu_L\\
\tau_L
\end{matrix}\right)^\dagger V^\dagger \left(\begin{matrix} \mathcal{O}(\epsilon^4) & \mathcal{O}(\epsilon^4) & 0\\
0 & y_\mu h_o\epsilon^2+\mathcal{O}(\epsilon^4) & 0\\
\mathcal{O}(\epsilon^4) & \mathcal{O}(\epsilon^4) & y_\tau h_o \epsilon+\mathcal{O}(\epsilon^3)
\end{matrix}\right)\left(\begin{matrix} e_R\\
\mu_R\\
\tau_R
\end{matrix}\right)
\end{equation}
where $\epsilon=\frac{m}{\Lambda}$. The matrix elements, $\mathcal{O}(\epsilon^3)$ and $\mathcal{O}(\epsilon^4)$, are  of the order of $\epsilon^3$ and $\epsilon^4$ respectively. They are the result of the higher order terms in Eq.~(\ref{eq:mclept}) containing cubic and quartic flavon products\footnotemark{}\addtocounter{footnote}{1}\addtocounter{Hfootnote}{1}. The mass matrix shown in Eq.~(\ref{eq:leptcontrib}) is approximately diagonalised~\footnote{The effect of higher order elements on diagonalisation is discussed in Section~4.} by left multiplying it with $V$. It is apparent that the charged-lepton masses, i.e.~the eigenvalues of the mass matrix, are in the ratio $\mathcal{O}(\epsilon):\mathcal{O}(\epsilon^2):\mathcal{O}(\epsilon^4)$. This is consistent with the experimentally-observed mass hierarchy, $\left(\frac{m_\mu}{m_e}\right) \approx \left(\frac{m_\tau}{m_\mu}\right)^2$.          

Now, we write the Dirac mass term for the neutrinos:
\begin{equation}\label{eq:mdl2dirac}
2 y_w L^\dagger \nu_R \tilde{H}
\end{equation}
where $\tilde{H}$ is the conjugate Higgs and $y_w$ is the coupling constant. With the help of Eq.~(\ref{eq:Tnu}), we also write the Majorana mass term for the neutrinos:
\begin{equation}\label{eq:mnurnur}
y_m \bar{S}_\nu^T\xi
\end{equation}
where $y_m$ is the coupling constant. Let $\langle\xi\rangle$ be the VEV acquired by the sextet flavon $\xi$, and let $\boldsymbol{ \langle\xi\rangle}$ be the corresponding $3\times 3$ symmetric matrix of the form given in Eq.~(\ref{eq:Tnu}). Combining the mass terms, Eq.~(\ref{eq:mdl2dirac}) and Eq.~(\ref{eq:mnurnur}), and using the VEVs of the Higgs and the flavon, we obtain the Dirac-Majorana mass matrix:
\begin{equation}
M=\left(\begin{matrix}0 & y_w h_o I\\
       y_w h_o I & \,\,y_m \boldsymbol{ \langle\xi\rangle}
\end{matrix}\right).
\end{equation}
The $6\times6$ mass matrix $M$, forms the coupling
\begin{equation}
M_{ij} \,\nu_i.\nu_j \quad \text{with} \quad  \nu=\left(\begin{matrix}\nu_{L}^*\\
	\nu_{R}
\end{matrix}\right)
\end{equation}
where $\nu_L=(\nu_e,\nu_\mu,\nu_\tau)^T$ are the left-handed neutrino flavour eigenstates.

Since $y_w h_o$ is at the electroweak scale and $y_m \boldsymbol{ \langle\xi\rangle}$ is at the high energy flavon scale ($>10^{10}$~GeV), small neutrino masses are generated through the see-saw mechanism. The resulting effective see-saw mass matrix is of the form
\begin{equation}\label{eq:seesaw}
M_\text{ss}=-\left(y_w h_o\right)^2\left(y_m {\boldsymbol{ \langle\xi\rangle}}\right)^{-1}.
\end{equation}
From Eq.~(\ref{eq:seesaw}), it is clear that the see-saw mechanism makes the light neutrino masses inversely proportional to the eigenvalues of the matrix $\boldsymbol{ \langle\xi\rangle}$. We now proceed to construct the four cases of the mass matrices, Eqs.~(\ref{eq:txmmat},~\ref{eq:tpmmat}), all of which result in the neutrino mass ratios, Eq.~(\ref{eq:numass}). To achieve this we choose suitable vacuum alignments\footnote{Refer to Appendix~C for the details of the flavon potentials that lead to these VEVs.} for the sextet flavon $\xi$. 

\subsection{$\txm_{(\chi=+\frac{\pi}{16})}$} 

Here we assign the vacuum alignment
\begin{equation}\label{eq:vevtxmp}
\langle\xi\rangle = \left((2-\sqrt{2}),1,0,0,1,0\right)^Tm.
\end{equation}
Using the symmetric matrix form of the sextet given in Eq.~(\ref{eq:Tnu}), we obtain
\begin{equation}\label{eq:vevsym1}
\boldsymbol{ \langle\xi\rangle} =\left(\begin{matrix}(2-\sqrt{2}) & 0 & \frac{1}{\sqrt{2}}\\
0 & 1 & 0\\
\frac{1}{\sqrt{2}} & 0 & 0
\end{matrix}\right)m.
\end{equation}
Diagonalising the corresponding effective see-saw mass matrix $M_{ss}$, Eq.~(\ref{eq:seesaw}), we get
\begin{equation}\label{eq:nucontrib}
U_\nu^\dagger M_{ss} U_\nu^* = \frac{\left(y_w h_o\right)^2}{y_m m} \text{Diag}\left({\textstyle\frac{\left(2+\sqrt{2}\right)}{1+\sqrt{2(2+\sqrt{2})}},1,\frac{\left(2+\sqrt{2}\right)}{-1+\sqrt{2(2+\sqrt{2})}}}\right)
\end{equation}
leading to the neutrino mass ratios, Eq.~(\ref{eq:numass}). The unitary matrix $U_\nu$ is given by
\begin{equation}\label{eq:unu1}
U_\nu = i \left(\begin{matrix}\cos\left(\frac{3\pi}{16}\right) & 0 & -i\sin\left(\frac{3\pi}{16}\right)\\
0 & 1 & 0\\
\sin\left(\frac{3\pi}{16}\right) & 0 & i\cos\left(\frac{3\pi}{16}\right)
\end{matrix}\right).
\end{equation}
The product of the contribution from the charged-lepton sector i.e.~$V$ from Eqs.~(\ref{eq:leptcontrib},~\ref{eq:gen3}) and the contribution from the neutrino sector i.e.~$U_\nu$ from Eqs.~(\ref{eq:nucontrib},~\ref{eq:unu1}) results in the $\txm_{(\chi=+\frac{\pi}{16})}$ mixing:
\begin{equation}\label{eq:mix1}
\begin{split}
&U_\text{PMNS}=VU_\nu=\\ &\left(\begin{matrix}1 & 0 & 0\\
0 & \om & 0\\
0 & 0 & \ob
\end{matrix}\right)\left(\begin{matrix}\sqrt{\frac{2}{3}}\cos \chi & \frac{1}{\sqrt{3}} & \sqrt{\frac{2}{3}}\sin \chi\\
-\frac{\cos \chi}{\sqrt{6}}-i\frac{\sin \chi}{\sqrt{2}} & \frac{1}{\sqrt{3}} & i\frac{\cos \chi}{\sqrt{2}}-\frac{\sin \chi}{\sqrt{6}}\\
-\frac{\cos \chi}{\sqrt{6}}+i\frac{\sin \chi}{\sqrt{2}} & \frac{1}{\sqrt{3}} & -i\frac{\cos \chi}{\sqrt{2}}-\frac{\sin \chi}{\sqrt{6}}
\end{matrix}\right)\left(\begin{matrix}1 & 0 & 0\\
0 & 1 & 0\\
0 & 0 & i
\end{matrix}\right)
\end{split}
\end{equation}
with $\chi=+\frac{\pi}{16}$.

\subsection{$\txm_{(\chi=-\frac{\pi}{16})}$} 
Here we assign the vacuum alignment
\begin{equation}\label{eq:vevtxmm}
\langle\xi\rangle = \left(0,1,(2-\sqrt{2}),0,1,0\right)^Tm
\end{equation}
resulting in the symmetric matrix
\begin{equation}\label{eq:vevsym2}
\boldsymbol{ \langle\xi\rangle}=\left(\begin{matrix}0 & 0 & \frac{1}{\sqrt{2}}\\
0 & 1 & 0\\
\frac{1}{\sqrt{2}} & 0 & (2-\sqrt{2})
\end{matrix}\right)m.
\end{equation}
In this case, the diagonalising matrix is
\begin{equation}
U_\nu =i \left(\begin{matrix}\cos\left(\frac{5\pi}{16}\right) & 0 & i\sin\left(\frac{5\pi}{16}\right)\\
0 & 1 & 0\\
\sin\left(\frac{5\pi}{16}\right) & 0 & -i\cos\left(\frac{5\pi}{16}\right)
\end{matrix}\right)
\end{equation}
and the corresponding mixing matrix is
\begin{equation}
\begin{split}
&U_\text{PMNS}=VU_\nu=\\&\left(\begin{matrix}1 & 0 & 0\\
0 & \om & 0\\
0 & 0 & \ob
\end{matrix}\right)\left(\begin{matrix}\sqrt{\frac{2}{3}}\cos \chi & \frac{1}{\sqrt{3}} & \sqrt{\frac{2}{3}}\sin \chi\\
-\frac{\cos \chi}{\sqrt{6}}-i\frac{\sin \chi}{\sqrt{2}} & \frac{1}{\sqrt{3}} & i\frac{\cos \chi}{\sqrt{2}}-\frac{\sin \chi}{\sqrt{6}}\\
-\frac{\cos \chi}{\sqrt{6}}+i\frac{\sin \chi}{\sqrt{2}} & \frac{1}{\sqrt{3}} & -i\frac{\cos \chi}{\sqrt{2}}-\frac{\sin \chi}{\sqrt{6}}
\end{matrix}\right)\left(\begin{matrix}1 & 0 & 0\\
0 & 1 & 0\\
0 & 0 & -i
\end{matrix}\right)\label{eq:mix2}
\end{split}
\end{equation}
with $\chi=-\frac{\pi}{16}$.

\subsection{$\tpm_{(\phi=+\frac{\pi}{16})}$} 
Here we assign the vacuum alignment
\begin{equation}\label{eq:vevtpmp}
\langle\xi\rangle = \left(i+\frac{1-i}{\sqrt{2}},1,-i+\frac{1+i}{\sqrt{2}},0, (\sqrt{2}-1), 0\right)^Tm
\end{equation}
resulting in the symmetric matrix
\begin{equation}\label{eq:vevsym3}
\boldsymbol{ \langle\xi\rangle}=\left(\begin{matrix}i+\frac{1-i}{\sqrt{2}} & 0 & 1-\frac{1}{\sqrt{2}}\\
0 & 1 & 0\\
1-\frac{1}{\sqrt{2}} & 0 & -i+\frac{1+i}{\sqrt{2}}
\end{matrix}\right)m.
\end{equation}
In this case, the diagonalising matrix is
\begin{equation}
U_\nu =i \left(\begin{matrix}\frac{1}{\sqrt{2}} e^{-i\frac{\pi}{16}} & 0 & -\frac{1}{\sqrt{2}} e^{-i\frac{\pi}{16}}\\
0 & 1 & 0\\
\frac{1}{\sqrt{2}} e^{i\frac{\pi}{16}} & 0 & \frac{1}{\sqrt{2}} e^{i\frac{\pi}{16}}
\end{matrix}\right)
\end{equation}
and the corresponding mixing matrix is
\begin{equation}\label{eq:mix3}
\begin{split}
&U_\text{PMNS}=VU_\nu=\\&\left(\begin{matrix}1 & 0 & 0\\
0 & \om & 0\\
0 & 0 & \ob
\end{matrix}\right)\left(\begin{matrix}\sqrt{\frac{2}{3}}\cos \phi & \frac{1}{\sqrt{3}} & \sqrt{\frac{2}{3}}\sin \phi\\
-\frac{\cos \phi}{\sqrt{6}}-\frac{\sin \phi}{\sqrt{2}} & \frac{1}{\sqrt{3}} & \frac{\cos \phi}{\sqrt{2}}-\frac{\sin \phi}{\sqrt{6}}\\
-\frac{\cos \phi}{\sqrt{6}}+\frac{\sin \phi}{\sqrt{2}} & \frac{1}{\sqrt{3}} & -\frac{\cos \phi}{\sqrt{2}}-\frac{\sin \phi}{\sqrt{6}}
\end{matrix}\right)\left(\begin{matrix}1 & 0 & 0\\
0 & 1 & 0\\
0 & 0 & i
\end{matrix}\right)
\end{split}
\end{equation}
with $\phi=+\frac{\pi}{16}$.

\subsection{$\tpm_{(\phi=-\frac{\pi}{16})}$} 
Here we assign the vacuum alignment
\begin{equation}\label{eq:vevtpmm}
\langle\xi\rangle =  \left(-i+\frac{1+i}{\sqrt{2}},1,i+\frac{1-i}{\sqrt{2}},0, (\sqrt{2}-1), 0\right)^Tm
\end{equation}
resulting in the symmetric matrix
\begin{equation}\label{eq:vevsym4}
\boldsymbol{ \langle\xi\rangle}=\left(\begin{matrix}-i+\frac{1+i}{\sqrt{2}} & 0 & 1-\frac{1}{\sqrt{2}}\\
0 & 1 & 0\\
1-\frac{1}{\sqrt{2}} & 0 & i+\frac{1-i}{\sqrt{2}}
\end{matrix}\right)m.
\end{equation}
In this case, the diagonalising matrix is
\begin{equation}
U_\nu =i \left(\begin{matrix}\frac{1}{\sqrt{2}} e^{i\frac{\pi}{16}} & 0 & \frac{1}{\sqrt{2}} e^{i\frac{\pi}{16}}\\
0 & 1 & 0\\
\frac{1}{\sqrt{2}} e^{-i\frac{\pi}{16}} & 0 & -\frac{1}{\sqrt{2}} e^{-i\frac{\pi}{16}}
\end{matrix}\right)
\end{equation}
and the corresponding mixing matrix is
\begin{equation}
\begin{split}
&U_\text{PMNS}=VU_\nu=\\&\left(\begin{matrix}1 & 0 & 0\\
0 & \om & 0\\
0 & 0 & \ob
\end{matrix}\right)\left(\begin{matrix}\sqrt{\frac{2}{3}}\cos \phi & \frac{1}{\sqrt{3}} & \sqrt{\frac{2}{3}}\sin \phi\\
-\frac{\cos \phi}{\sqrt{6}}-\frac{\sin \phi}{\sqrt{2}} & \frac{1}{\sqrt{3}} & \frac{\cos \phi}{\sqrt{2}}-\frac{\sin \phi}{\sqrt{6}}\\
-\frac{\cos \phi}{\sqrt{6}}+\frac{\sin \phi}{\sqrt{2}} & \frac{1}{\sqrt{3}} & -\frac{\cos \phi}{\sqrt{2}}-\frac{\sin \phi}{\sqrt{6}}
\end{matrix}\right)\left(\begin{matrix}1 & 0 & 0\\
0 & 1 & 0\\
0 & 0 & -i
\end{matrix}\right)\label{eq:mix4}
\end{split}
\end{equation}
with $\phi=-\frac{\pi}{16}$.

As stated earlier, the four cases, Eqs.~(\ref{eq:vevsym1},~\ref{eq:vevsym2},~\ref{eq:vevsym3},~\ref{eq:vevsym4}), result in the same neutrino mass ratios, Eq.~(\ref{eq:numass}).

\subsection*{Symmetries of the VEVs of the sextet flavons}

A careful inspection of the Majorana matrices, Eqs.~(\ref{eq:txmmat},~\ref{eq:tpmmat}), reveals several symmetries which could be attributed to the underlying symmetries of the VEVs of the sextet flavons, Eqs.~(\ref{eq:vevtxmp},~\ref{eq:vevtxmm},~\ref{eq:vevtpmp},~\ref{eq:vevtpmm}). The VEVs, Eqs.~(\ref{eq:vevtxmp},~\ref{eq:vevtxmm}), (and thus the mass matrices, Eqs.~(\ref{eq:txmmat})) are composed of real numbers implying they remain invariant under complex conjugation. Therefore, they do not contribute to $CP$ violation. In our model, $U_\text{PMNS}=V U_\nu$ where $V$ originates from the charged-lepton mass matrix, Eq.~(\ref{eq:leptcontrib}). Since $V$ is maximally $CP$-violating ($\delta=\frac{\pi}{2}$), the resulting leptonic mixing, $V U_\nu$, is also maximally $CP$-violating ($\txm$). Note that $U_{\txm}$, Eq.~(\ref{eq:txmform}), is symmetric under the conjugation and the exchange of $\mu$ and $\tau$ rows. This generalised $CP$ symmetry under the combined operations of $\mu\text{-}\tau$ exchange and complex conjugation is referred to as $\mu\text{-}\tau$ reflection symmetry in previous publications~\cite{Harrison02, Feruglio13, Harrison02c, Harrison04, Grimus04}. The conjugation symmetry in the neutrino VEVs together with maximal $CP$ violation from the charged-lepton sector produces the $\mu\text{-}\tau$ reflection symmetry of $U_\text{PMNS}$.

Consider the exchange of the first and the third rows as well as the columns of the mass matrix, Eq.~(\ref{eq:Tnu}). This is equivalent to the exchange of the first and the third elements and the fourth and the sixth elements of the sextet flavon, Eq.~(\ref{eq:xinu}). In $\Sigma(72\times3)$, this exchange can be realised using the group transformation by the unitary matrix $E.V.V$,
\begin{equation}\label{eq:res1}
E.V.V\equiv
\left(\begin{matrix}0 & 0 & -1\\
       0 & -1 & 0\\
       -1 & 0 & 0
\end{matrix}\right)\equiv
\left(\begin{matrix}0 & 0 & 1 & 0 & 0 & 0\\
	0 & 1 & 0 & 0 & 0 & 0\\
	1 & 0 & 0 & 0 & 0 & 0\\
	0 & 0 & 0 & 0 & 0 & 1\\
	0 & 0 & 0 & 0 & 1 & 0\\
	0 & 0 & 0 & 1 & 0 & 0
\end{matrix}\right),
\end{equation}
with $E$ and $V$ given in Eqs.~(\ref{eq:gen3}, \ref{eq:gen6}). By the group transformation we imply left and right multiplication of the mass matrix using the $3\times 3$ unitary matrix and its transpose or equivalently left multiplication of the sextet flavon using the $6\times 6$ unitary matrix. The mass matrices, Eqs.~(\ref{eq:tpmmat}), and the corresponding flavon VEVs, Eqs.~(\ref{eq:vevtpmp},~\ref{eq:vevtpmm}), are invariant under the transformation by $E.V.V$ together with the conjugation. The VEVs break $\Sigma(72\times3)$ almost completely except for $E.V.V$ with conjugation which remains as their residual symmetry\footnote{Here we apply $E.V.V$ and complex conjugation together even though complex conjugation is not a part of $\Sigma(72\times3)$.}. The resulting mixing matrix, $U_\text{PMNS}=V U_\nu$, is $\tpm$ which is real and $CP$ conserving. $E.V.V$-conjugation symmetry in the neutrino VEVs together with maximal $CP$ violation from the charged-lepton sector produces the $CP$ symmetry of $U_\text{PMNS}$.

\begin{comment}
 In our model, the second column becomes trimaximal because of the vanishing of the forth and the sixth elements of the VEVs, Eqs.~(\ref{eq:vevtxmp},~\ref{eq:vevtxmm},~\ref{eq:vevtpmp},~\ref{eq:vevtpmm}) which correspond to the off-diagonal zeros present in the mass matrices, Eqs.~(\ref{eq:txmmat},~\ref{eq:tpmmat}).
\end{comment}
\begin{comment}
\begin{equation}\label{eq:res2}
E.E.A.E\equiv
\left(\begin{matrix}-1 & 0 & 0\\
       0 & 1 & 0\\
       0 & 0 & -1
\end{matrix}\right)
\end{equation}
Another interesting transformation is the group action by the unitary matrix 
\begin{equation}\label{eq:res3}
E.E.A(16).E\equiv
\left(\begin{matrix}-1 & 0 & 0\\
       0 & 1 & 0\\
       0 & 0 & -1
\end{matrix}\right)
\end{equation}
where $A(16)$ is diag$(1,e^\frac{2\pi i}{16},e^\frac{-2\pi i}{16})$ as defined in Ref.~\cite{Sigma1}
\end{comment}

Both $\txm$ and $\tpm$ have a trimaximal second column. This feature of the mixing matrix was linked to the "magic" symmetry of the mass matrix~\cite{Harrison04, Friedberg06, Lam06, Luo07}. In our model, the charged-leptonic contribution, $V$, is trimaximal. Because of the vanishing of the forth and the sixth elements of the sextet VEVs, Eqs.~(\ref{eq:vevtxmp},~\ref{eq:vevtxmm},~\ref{eq:vevtpmp},~\ref{eq:vevtpmm}), which correspond to the off-diagonal (1-2, 2-3) zeros present in the mass matrices, Eqs.~(\ref{eq:txmmat},~\ref{eq:tpmmat}), the trimaximality of $V$ carries over to $U_\text{PMNS}=V U_\nu$. Consider the unitary matrix,
\begin{equation}\label{eq:res2}
A\equiv\text{diag}(-1,1,-1).
\end{equation}
Group transformation by $A$ results in the multiplication of the off-diagonal (1-2, 2-3) elements of the Majorana matrix by $-1$. Invariance under $A$, implies these elements vanish and ensures trimaximality. In the literature, small groups like the Klein group~\cite{Feruglio13, Toorop11, Lam07, Lam08, Lam08b, Lam11} are often used to implement symmetries like the generalised $CP$ and the trimaximality as the residual symmetries of the mass matrix. However, $A$ is not a group member of $\Sigma(72\times3)$. In our model, the vanishing mass matrix elements arise as a consequence of the specific choice of the flavon potential, Eq.~(\ref{eq:potxi}), rather than the result of a residual symmetry under $\Sigma(72\times3)$. 

The presence of a simple set of numbers in the VEVs (and the mass matrices) is suggestive of additional symmetry transformations (like the one generated by $A$, Eq.~(\ref{eq:res2})) which are not a part of $\Sigma(72\times3)$. The present model only serves as a template for constructing any fully constrained Majorana mass matrix using $\Sigma(72\times3)$. We impose additional symmetries on the mass matrix by using flavon potentials with a carefully chosen set of parameters, Table~\ref{tab:constants}. Realising these symmetries naturally by incorporating more group transformations along with $\Sigma(72\times3)$ in an expanded flavour group requires further investigation.

%----------------------------------------------------------------------------------------------------------------------------------
\section{Predicted Observables}

For comparing our model with the neutrino oscillation experimental data, we use the global analysis done by the NuFIT group and their latest results reproduced in Eqs.~(\ref{eq:anglevalues1}-\ref{eq:massvalues2}). They are a leading group doing a comprehensive statistical data analysis based on essentially all currently available neutrino oscillation experiments. Their results are updated regularly and published on the NuFIT website~\cite{Esteban17}. The value $\sin^2 \theta_{13} = \frac{2}{3} \sin^2 \frac{\pi}{16} = 0.02537$,\footnote{Besides in Ref.~\cite{LIS}, this value was predicted in the context of $\Delta(6n^2)$ symmetry group in Ref.~\cite{Delta16} and later obtained in Ref.~\cite{Thomas}} is slightly more than the upper limit of the $3\sigma$ range, $0.02392$. We provide a solution to this discrepancy in the following discussion.

In our previous analysis in Subsections~3.1-3.4, we used the relation $U_\text{PMNS}=VU_\nu$ where $V$ is the left-diagonalising matrix for the charged-lepton mass matrix, Eq.~(\ref{eq:leptcontrib}). However, the diagonalisation achieved by $V$ is only an approximation. In Eq.~(\ref{eq:leptcontrib}), the presence of the ${\mathcal O}(\epsilon^4)$ element in the $e_L\text{-}\mu_R$ off-diagonal position in relation to the ${\mathcal O}(\epsilon^4)$ electron mass and ${\mathcal O}(\epsilon^2)$ muon mass produces an ${\mathcal O}(\epsilon^2)$ correction to the diagonalisation, i.e. a more accurate left-diagonalisation matrix is
\begin{equation}
\left(\begin{matrix}\mathcal{O}(1) & \mathcal{O}(\epsilon^2) & 0 \\
\mathcal{O}(\epsilon^2) & \mathcal{O}(1) & 0\\
0 & 0 & 1
\end{matrix}\right).V.
\end{equation}
The resulting correction in the $e3$ element of $U_\text{PMNS}$ is
\begin{equation}
(U_\text{PMNS})_{e3}\rightarrow (U_\text{PMNS})_{e3}+\mathcal{O}(\epsilon^2) ( U_\text{PMNS})_{\mu3}.
\end{equation}
Since $(U_\text{PMNS})_{e3}=\sin \theta_{13} e^{-i\delta}$ and $\epsilon\approx\frac{m_\mu}{m_\tau}=\mathcal{O}(0.1)$, we obtain
\begin{equation}
\sin \theta_{13} e^{-i\delta} \rightarrow \sin \theta_{13} e^{-i\delta} + \mathcal{O}(0.01).
\end{equation}
The above correction is sufficient to reduce\footnote{Whether this correction has a reducing or enhancing effect on $\sin^2 \theta_{13}$, is determined by the relative phase between the $(U_\text{PMNS})_{e3}$ and the correction, which in turn is determined by the phases of the elements in the mass matrix, Eq.~(\ref{eq:leptcontrib}). For a range of values of the mass matrix elements, we have numerically verified that a reducing effect can be achieved.} our prediction for $\sin^2 \theta_{13}$ to within the $3\sigma$ range. 
%=\Delta(\sin \theta_{13} e^{-i\delta}) 

For the solar angle, using the formula given in Table~\ref{tab:anglesandphase}, we get
\begin{equation}\label{eq:theta12}
\begin{split}
\sin^2 \theta_{12} &= \frac{1}{3-2\sin^2\left(\frac{\pi}{16}\right)}\\ 
&= 0.342\,.
\end{split}
\end{equation}
This is within $3\sigma$ errors of the experimental values, although there is a small tension towards the upper limit. For the atmospheric angle, $\txm$ predicts maximal mixing:
\begin{equation}\label{eq:theta23x}
\sin^2 \theta_{23} = \frac{1}{2}\,
\end{equation}
which is also within $3\sigma$ errors. The NuFIT data as well as other global fits~\cite{Capozzi17,Capozzi16} are showing a preference for non-maximal atmospheric mixing. As a result there has been a lot of interest in the problem of octant degeneracy of $\theta_{23}$~\cite{Agarwalla13,Chatterjee13,Choubey13,Das15,Agarwalla16,Choubey16,Bora16,Chatterjee16}. $\tpm$ predicts this non-maximal scenario of atmospheric mixing. $\tpm_{(\phi=\frac{\pi}{16})}$ and $\tpm_{(\phi=-\frac{\pi}{16})}$ correspond to the first and the second octant solutions respectively. Using the formula for $\theta_{23}$ given in Table~\ref{tab:anglesandphase}, we get 
\begin{gather}
\tpm_{(\phi=+\frac{\pi}{16})}:\,
\begin{split}\label{eq:theta23p1}
\quad \sin^2 \theta_{23} &= \frac{2\sin^2\left(\frac{2\pi}{3}+\frac{\pi}{16}\right)}{3-2\sin^2\left(\frac{\pi}{16}\right)} \\
&= 0.387\,,\\
\end{split}\\
%\notag\\
\tpm_{(\phi=-\frac{\pi}{16})}:\,
\begin{split}\label{eq:theta23p2}
\quad \sin^2 \theta_{23} &= \frac{2\sin^2\left(\frac{2\pi}{3}-\frac{\pi}{16}\right)}{3-2\sin^2\left(\frac{\pi}{16}\right)}\\
&= 0.613\,.
\end{split}
\end{gather}
The Dirac $CP$ phase, $\delta$, has not been measured yet. The discovery that the reactor mixing angle is not very small has raised the possibility of a relatively earlier measurement of $\delta$~\cite{Ohlsson13,Agarwalla14,Girardi15}. $\txm$ having $\delta=\pm\frac{\pi}{2}$ should lead to large observable $CP$-violating effects. Substituting $\chi=\pm\frac{\pi}{16}$ in Eq.~(\ref{eq:jcp}), our model gives
\begin{equation}
\begin{split}
J&=\pm\frac{\sin \frac{\pi}{8}}{6\sqrt{3}}\\
&= \pm0.0368
\end{split}
\end{equation} 
which is about $40\%$ of the maximum value of the theoretical range, $-\frac{1}{6\sqrt{3}}\leq J\leq +\frac{1}{6\sqrt{3}}$.
On the other hand, $\tpm$, with $\delta=0,\,\pi$ and $J=0$, is $CP$ conserving. 

\begin{figure}[]
\begin{center}
\includegraphics[scale=0.7]{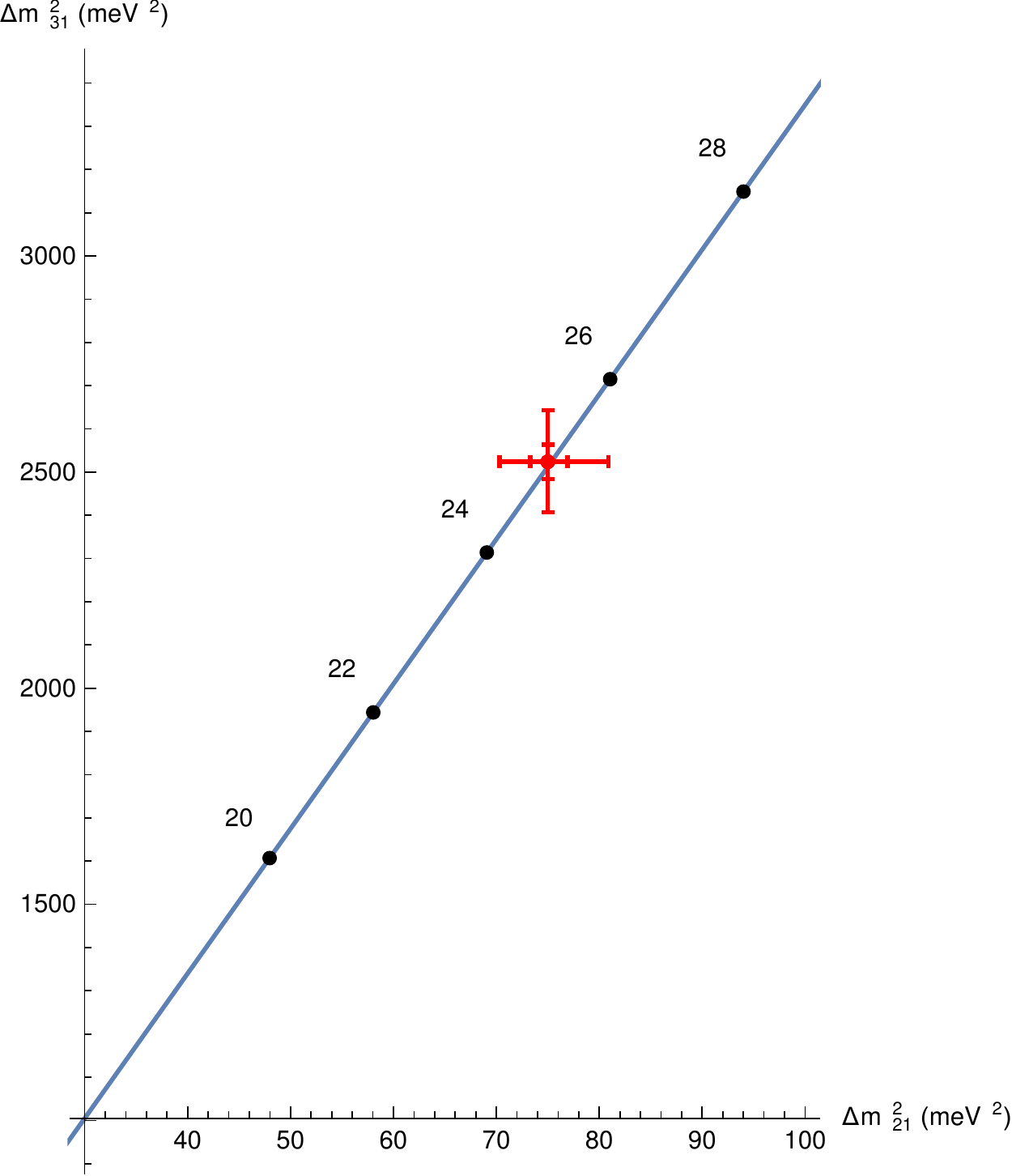}
\caption[$\Delta m_{31}^2$ vs $\Delta m_{21}^2$ plane]{$\Delta m_{31}^2$ vs $\Delta m_{21}^2$ plane. The straight line shows the neutrino mass ratios Eq.~(\ref{eq:numass}). As a parametric plot, the line can be represented as $\Delta m_{21}^2=(r_{21}^2-1)m_1^2$ and $\Delta m_{31}^2=(r_{31}^2-1)m_1^2$ where $r_{21}=\frac{m_2}{m_1}=\frac{1+\sqrt{2(2+\sqrt{2})}}{\left(2+\sqrt{2}\right)}$ and $r_{31}=\frac{m_3}{m_1}=\frac{1+\sqrt{2(2+\sqrt{2})}}{-1+\sqrt{2(2+\sqrt{2})}}$ are the mass ratios obtained from Eq.~(\ref{eq:numass}). The parametric values of the light neutrino mass, $m_1$, (denoted by the black dots on the line) are in terms of meV. The red marking indicates the experimental best fit for $\Delta m_{21}^2$ and $\Delta m_{31}^2$ along with $1\sigma$ and $3\sigma$ errors.}
\label{fig:neutrinopredict2}
\end{center}
\end{figure}

The neutrino mixing angles are fully determined by the model, Eqs.~(\ref{eq:theta13}, \ref{eq:theta12}, \ref{eq:theta23x}, \ref{eq:theta23p1}, \ref{eq:theta23p2}). Hence, we simply compared the individual mixing angles with the experimental data in the earlier part of this section. Regarding the neutrino masses, the model predicts their ratios, Eq.~(\ref{eq:numass}). To compare this result with the experimental data, which gives the mass-squared differences, Eqs.~(\ref{eq:massvalues1},~\ref{eq:massvalues2}), we utilise a $\chi^2$ analysis, 
\begin{equation}\label{eq:X2}
\chi^2=\displaystyle\sum_{{\displaystyle x}=\Delta m^2_{21}, \Delta m^2_{31}} \left(\frac{x_\text{model}-x_\text{expt}}{\sigma_{x\,\text{expt}}}\right)^2.
\end{equation}
We report that the predicted neutrino mass ratios are consistent with the experimental mass-squared differences. Using the $\chi^2$ analysis we obtain,
\begin{gather}
m_1=25.04^{+0.17}_{-0.15}~\text{meV},\notag\\
m_2=26.50^{+0.18}_{-0.16}~\text{meV},\label{eq:m123}\\
m_3=56.09^{+0.37}_{-0.34}~\text{meV}.\notag
\end{gather}
The best fit values correspond to $\chi^2_\text{min}=0.03$ and the error ranges correspond to $\Delta\chi^2=1$, where $\Delta\chi^2=\chi^2-\chi^2_\text{min}$. The results from our analysis are also shown in Figure~\ref{fig:neutrinopredict2}.

Note that the mass ratios Eq.~(\ref{eq:numass}), are incompatible with the inverted mass hierarchy. Considerable experimental studies are being conducted to determine the mass hierarchy~\cite{Agarwalla14,Ghosh13,Capozzi14,Winter13,Wang17,Simpson17,Rahaman17,Stanco17} and we may expect a resolution in the not-too-distant future. Observation of the inverted hierarchy will obviously rule out the model.

Cosmological observations can provide limits on the sum of the neutrino masses. The strongest such limit has been set recently by the data collected using the Planck satellite~\cite{Ade16,Giusarma16}:
\begin{equation}
\sum_i m_i < 183~\text{meV}.
\end{equation}
Our predictions Eqs.~(\ref{eq:m123}), give a sum 
\begin{equation}\label{eq:mcosmo}
\sum_i m_i = 107.6^{+0.71}_{-0.65}~\text{meV}
\end{equation}
which is not far below the current cosmological limit. Improvements in the cosmological bounds from Planck data are expected. Future ground-based CMB polarisation experiments such as Polarbear-2~\cite{Inoue16} and Square Kilometer Array-2~\cite{Oyama16}, could lower the cosmological limit to below $100~\text{meV}$ and could also determine the mass hierarchy. Such results may support or rule out our model.

Neutrinoless double beta decay experiments seek to determine the nature of the neutrinos as Majorana or not. These experiments have so far set limits on the effective electron neutrino mass~\cite{DBD} $|m_{\beta\beta}|$, where
\begin{align}
\begin{split}
m_{\beta\beta}&=m_1 U_{e1}^2+m_2 U_{e2}^2+m_3 U_{e3}^2\\
&=m_1 |U_{e1}|^2+m_2 |U_{e2}|^2 e^{i\alpha_{21}}+m_3 |U_{e3}|^2 e^{i(\alpha_{31}-2\delta)}
\end{split}
\end{align}
with $U$ representing $U_\text{PMNS}$. In all the four mixing scenarios predicted by the model, Eqs.~(\ref{eq:mix1},~\ref{eq:mix2},~\ref{eq:mix3},~\ref{eq:mix4}), we have $|U_{e1}|=\sqrt{\frac{2}{3}}\cos \frac{\pi}{16}$, $|U_{e2}|=\frac{1}{\sqrt{3}}$ and $|U_{e3}|=\sqrt{\frac{2}{3}}\sin \frac{\pi}{16}$. By comparing with the standard PDG parameterisation, Eq.~(\ref{eq:pmns}), we can also show that, all these scenarios lead to
\begin{equation}
\alpha_{21}=0, \quad \alpha_{31}-2\delta= \pi.
\end{equation}
Therefore the model predicts
\begin{equation}\label{eq:modelbb}
m_{\beta\beta}=\frac{2}{3}m_1\cos^2 \frac{\pi}{16}+\frac{1}{3}m_2-\frac{2}{3}m_3\sin^2 \frac{\pi}{16}.  
\end{equation}
Substituting the neutrino masses from Eqs.~(\ref{eq:m123}) in Eq.~(\ref{eq:modelbb}) we get 
\begin{equation}\label{eq:mbb}
m_{\beta\beta}=23.47^{+0.16}_{-0.14}~\text{meV}.
\end{equation}
The most stringent upper bounds on the value of $|m_{\beta\beta}|$ have been set by Heidelberg-Moscow~\cite{DBHM1,DBHM2}, Cuoricino~\cite{DBCouricino}, NEMO3~\cite{Gomez16}, EXO200~\cite{Albert14} and GERDA~\cite{DBGerda1} experiments. Combining their results leads to the bounds of the order of a few hundreds of meV~\cite{Guzowski16}. New experiments such as CUORE~\cite{Artusa15}, SuperNEMO~\cite{Vilela16} and GERDA-2~\cite{Andrea16} will improve the measurements on $|m_{\beta\beta}|$ to a few tens of meV and thus may support or rule out our model.

\subsection*{Renormalisation Effects on the Observables}

The see-saw mechanism requires the existence of a heavy Majorana mass term coupling the right-handed neuntrinos together. Our model, combined with the observed neutrino mass-squared differences, predicts that the neutrino masses are a few tens of meVs. This places the see-saw scale (also the flavon scale) at around $10^{12}$~GeV. As such, this is the scale at which the fully constrained mass matrices, as proposed in our model, are generated. In order to accurately compare the the model with the observed masses and mixing parameters, it is necessary to calculate its renormalisation group (RG) evolution from the high energy scale down to the electroweak scale. 

We use the Mathematica package, REAP~\cite{Antusch05}, to numerically study the RG evolution of the masses and the mixing observables. The Mathematica code for calculating the RG evolution relevant to the model is given below:
\begin{lstlisting}
Needs["REAP`RGESM`"];
RGEAdd["SM"];
RGESetInitial[10^12, 
 RGE\[Theta]12 -> 33.79 Degree, 
 RGE\[Theta]23 -> 45.00 Degree, 
 RGE\[Theta]13 -> 9.165 Degree, 
 RGE\[Delta] -> 90.00 Degree, 
 RGEMlightest -> 0.03055, 
 RGE\[CapitalDelta]m2sol -> 0.0001119, 
 RGE\[CapitalDelta]m2atm -> 0.0037490, 
 RGEY\[Nu] -> {{.01, 0, 0}, {0, .01, 0}, 
 {0, 0, .01}}];
RGESolve[100, 10^12];
MNSParameters[RGEGetSolution[100,
 RGEM\[Nu]], RGEGetSolution[100,RGEYe]]
\end{lstlisting}

In the above code, the initial values of the mixing observables and the masses are set at $10^{12}$~GeV. The mixing observables are chosen such that they correspond to $\txm_{(\chi=+\frac{\pi}{16})}$.  We set the masses to be $30.55$~meV, $32.33$~meV and $69.24$~meV. These specific values are chosen such that they are consistent with Eq.~(\ref{eq:numass}) (at $10^{12}$~GeV) and give the best fit to the observed mass-squared differences when renormalised to the electroweak scale ($100$~GeV). MSNParameters in the code gives the renormalised parameters at $100$~GeV as its output. Here we obtain $\theta_{12}=33.78^\circ$, $\theta_{23}=45.00^\circ$, $\theta_{13}=9.165^\circ$, $\delta=90.00^\circ$, $m_1=24.63$~meV, $m_2=26.07$~meV, $m_3=56.16$~meV~\footnote{These masses correspond to $\chi^2_\text{min}=1.23$ as calculated using Eq.~(\ref{eq:X2}).} as the output. From these values we conclude that, under the conditions of our model, renormalisation has virtually no effect on the mixing parameters. On the other hand, it affects our predictions for the masses, Eqs.~(\ref{eq:m123}, \ref{eq:mcosmo}, \ref{eq:mbb}), by a few percentage points.

Analysis of RG equations~\cite{Antusch05, Antusch03, Chankowski00, Casas00, Lola00, Harrison10} show that, even though the neutrino masses (the fermion masses in general) evolve appreciably, their ratios evolve slowly. This behaviour is sometimes referred to as "universal scaling". For our model, the light neutrino masses evolve by around $20\%$, while their ratios by less than $1\%$. This ensures that the mass ratios, Eq.~(\ref{eq:numass}), theorised at the high energy scale remain practically valid at the electroweak scale as well. 

\section{Summary}

In this paper we utilise the group $\Sigma(72\times3)$ to construct fully-constrained Majorana mass matrices for the neutrinos. These mass matrices reproduce the results obtained in Ref.~\cite{LIS} i.e.~$\txm_{(\chi=\pm\frac{\pi}{16})}$ and $\tpm_{(\phi=\pm\frac{\pi}{16})}$ mixings along with the neutrino mass ratios, Eq.~(\ref{eq:numass}). The mixing observables as well as the neutrino mass ratios are shown to be consistent with the experimental data. $\txm_{(\chi=\pm\frac{\pi}{16})}$ and $\tpm_{(\phi=\pm\frac{\pi}{16})}$ predict the Dirac $CP$-violating effect to be maximal (at fixed $\theta_{13}$) and null respectively. Using the neutrino mass ratios in conjunction with the experimentally-observed neutrino mass-squared differences, we calculate the individual neutrino masses. We note that our predicted neutrino mass ratios are incompatible with the inverted mass hierarchy. We also predict the effective electron neutrino mass for the neutrinoless double beta decay, $|m_{\beta\beta}|$. We briefly discuss the current status and future prospects of determining experimentally the neutrino observables leading to the confirmation or the falsification of our model. In the context of model-building, we carry out an in-depth analysis of the representations of $\Sigma(72\times3)$ and develop the necessary groundwork to construct the flavon potentials satisfying the $\Sigma(72\times3)$ flavour symmetry. In the charged-lepton sector, we use two triplet flavons with a suitably chosen set of VEVs which provide a $3\times 3$ trimaximal contribution towards the PMNS mixing matrix. It also explains the hierarchical structure of the charged lepton masses. In the neutrino sector, we discuss four cases of Majorana mass matrices. The $\Sigma(72\times3)$ sextet acts as the most general placeholder for a fully constrained Majorana mass matrix. The intended mass matrices are obtained by assigning appropriate VEVs to the sextet flavon. It should be noted that we need additional symmetries to `explain' any specific texture in the mass matrix.

This work was supported by the UK Science and Technology Facilities Council (STFC). Two of us (RK and PFH) acknowledge the hospitality of the Centre for Fundamental Physics (CfFP) at the Rutherford Appleton Laboratory. RK acknowledges the support from the University of Warwick. RK thanks the management of the School of the Good Shepherd, Thiruvananthapuram, for providing a convenient and flexible working arrangement conducive to research.
%----------------------------------------------------------------------------------------------------------------------------------
\section*{Appendix A:\quad Irreps of $\Sigma(72\times3)$ and their Tensor Product Expansions}

\begin{flalign}\label{eq:tensor33}
&\text{{\textit {\textbf {i}}})\,\,\,}\boldsymbol{3}\otimes\boldsymbol{3}=\boldsymbol{6}\oplus\boldsymbol{\tb}&
\end{flalign}
The generator matrices for the triplet representation are provided in Eq.~(\ref{eq:gen3}). We define the basis for the sextet representation using Eqs.~(\ref{eq:tensor1exp}). The resulting generator matrices are 
\begin{align}\label{eq:gen6}
\begin{split}
&C \equiv
\left(\begin{matrix}1 & 0 & 0 & 0 & 0 & 0\\
	0 & \ob & 0 & 0 & 0 & 0\\
	0 & 0 & \om & 0 & 0 & 0\\
	0 & 0 & 0 & 1 & 0 & 0\\
	0 & 0 & 0 & 0 & \ob & 0\\
	0 & 0 & 0 & 0 & 0 & \om
\end{matrix}\right), \quad E \equiv
\left(\begin{matrix}0 & 1 & 0 & 0 & 0 & 0\\
	0 & 0 & 1 & 0 & 0 & 0\\
	1 & 0 & 0 & 0 & 0 & 0\\
	0 & 0 & 0 & 0 & 1 & 0\\
	0 & 0 & 0 & 0 & 0 & 1\\
	0 & 0 & 0 & 1 & 0 & 0
\end{matrix}\right),\\
&V\equiv
-\frac{1}{3}\left(\begin{matrix}1 & 1 & 1 & \sqrt{2} & \sqrt{2} & \sqrt{2}\\
	1 & \ob & \om & \sqrt{2} & \sqrt{2}\ob & \sqrt{2}\om\\
	1 & \om & \ob & \sqrt{2} & \sqrt{2}\om & \sqrt{2}\ob\\
	\sqrt{2} & \sqrt{2} & \sqrt{2} & -1 & -1 & -1\\
	\sqrt{2} & \sqrt{2}\ob & \sqrt{2}\om & -1 & -\ob & -\om\\
	\sqrt{2} & \sqrt{2}\om & \sqrt{2}\ob & -1 & -\om & -\ob
\end{matrix}\right),\\
&X\equiv
-\frac{1}{3}\left(\begin{matrix}1 & 1 & \om & \sqrt{2}\ob & \sqrt{2}\ob & \sqrt{2}\\
	1 & \ob & \ob & \sqrt{2}\ob & \sqrt{2}\om & \sqrt{2}\om\\
	\ob & 1 & \ob & \sqrt{2}\om & \sqrt{2}\ob &  \sqrt{2}\om\\
	 \sqrt{2}\om & \sqrt{2}\om &  \sqrt{2}\ob & -1 & -1 & -\om\\
	 \sqrt{2}\om &  \sqrt{2} &  \sqrt{2} & -1 & -\ob & -\ob\\
	 \sqrt{2} &  \sqrt{2}\om &  \sqrt{2} & -\ob & -1 & -\ob
\end{matrix}\right).
\end{split}
\end{align}

\begin{flalign}\label{eq:tensor33b}
&\text{{\textit {\textbf {ii}}})\,\,\,}\boldsymbol{3}\otimes\boldsymbol{\tb}=\boldsymbol{1}\oplus\boldsymbol{8}&
\end{flalign}
With $(a_1,a_2,a_3)^T$ and $(\B_1,\B_2,\B_3)^T$ transforming as $\boldsymbol{3}$ and $\boldsymbol{\tb}$, the tensor product expansion, Eq.~(\ref{eq:tensor33b}), is given by
\begin{align}\label{eq:cg33b}
\begin{split}
&\boldsymbol{1}\equiv \frac{1}{\sqrt{3}}\left(a_1 \B_1 + a_2 \B_2+ a_3 \B_3\right),\\
&\boldsymbol{8} \equiv
\left(\begin{matrix}\frac{1}{\sqrt{6}}a_1 \bar{b}_1 - \frac{\sqrt{2}}{\sqrt{3}}a_2 \bar{b}_2+\frac{1}{\sqrt{6}}a_3 \bar{b}_3\\
	\frac{1}{\sqrt{2}}\left(a_1 \B_1 - a_3 \B_3\right)\\
	\frac{1}{\sqrt{2}}\left(a_2 \B_3 + a_3 \B_2\right)\\
	\frac{1}{\sqrt{2}}\left(a_3 \B_1 + a_1 \B_3\right)\\
	\frac{1}{\sqrt{2}}\left(a_1 \B_2 + a_2 \B_1\right)\\
	\frac{i}{\sqrt{2}}\left(a_2 \B_3 - a_3 \B_2\right)\\
	\frac{i}{\sqrt{2}}\left(a_3 \B_1 - a_1 \B_3\right)\\
      	\frac{i}{\sqrt{2}} \left(a_1 \B_2 - a_2 \B_1\right)
\end{matrix}\right).
\end{split}
\end{align}
In this basis, the generator matrices of the octet representation are 
\begin{align}\label{eq:gen8}
\begin{split}
&C \equiv
\frac{1}{2}\left(\begin{matrix}2 & 0 & 0 & 0 & 0 & 0 & 0 & 0\\
	0 & 2 & 0 & 0 & 0 & 0 & 0 & 0\\
	0 & 0 & -1 & 0 & 0 & -\sqrt{3} & 0 & 0\\
	0 & 0 & 0 & -1 & 0 & 0 & -\sqrt{3} & 0\\
	0 & 0 & 0 & 0 & -1 & 0 & 0 & -\sqrt{3}\\
	0 & 0 & \sqrt{3} & 0 & 0 & -1 & 0 & 0\\
	0 & 0 & 0 & \sqrt{3} & 0 & 0 & -1 & 0\\
	0 & 0 & 0 & 0 & \sqrt{3} & 0 & 0 & -1
\end{matrix}\right),\\
&E \equiv
\frac{1}{2}\left(\begin{matrix}-1 & \sqrt{3} & 0 & 0 & 0 & 0 & 0 & 0\\
	-\sqrt{3} & -1 & 0 & 0 & 0 & 0 & 0 & 0\\
	0 & 0 & 0 & 2 & 0 & 0 & 0 & 0\\
	0 & 0 & 0 & 0 & 2 & 0 & 0 & 0\\
	0 & 0 & 2 & 0 & 0 & 0 & 0 & 0\\
	0 & 0 & 0 & 0 & 0 & 0 & 2 & 0\\
	0 & 0 & 0 & 0 & 0 & 0 & 0 & 2\\
	0 & 0 & 0 & 0 & 0 & 2 & 0 & 0
\end{matrix}\right),\\
&V\equiv
\frac{1}{6}\left(\begin{matrix}0 & 0 &  \sqrt{3} &  \sqrt{3} &  \sqrt{3} & 3 & 3 & 3\\
	0 & 0 & 3 & 3 & 3 & - \sqrt{3} & - \sqrt{3} & - \sqrt{3}\\
	 \sqrt{3} & 3 & 4 & -2 & -2 & 0 & 0 & 0\\
	 \sqrt{3} & 3 & -2 & 1 & 1 & -2 \sqrt{3} &  \sqrt{3} &  \sqrt{3}\\
	 \sqrt{3} & 3 & -2 & 1 & 1 & 2 \sqrt{3} & - \sqrt{3} & - \sqrt{3}\\
	-3 &  \sqrt{3} & 0 & 2 \sqrt{3} & -2 \sqrt{3} & 0 & 0 & 0\\
	-3 &  \sqrt{3} & 0 & - \sqrt{3} &  \sqrt{3} & 0 & -3 & 3\\
	-3 &  \sqrt{3} & 0 & - \sqrt{3} &  \sqrt{3} & 0 & 3 & -3
\end{matrix}\right),\\
&X\equiv
\frac{1}{6}\left(\begin{matrix}0 & 0 &  -2\sqrt{3} &  \sqrt{3} &  \sqrt{3} & 0 & -3 & 3\\
	0 & 0 & 0 & -3 & 3 & 2\sqrt{3} & - \sqrt{3} & - \sqrt{3}\\
	 \sqrt{3} & -3 & 1 & 1 & 4 & -\sqrt{3} & \sqrt{3} & 0\\
	 -2\sqrt{3} & 0 & -2 & -2 & 1 & 0 &  2\sqrt{3} &  \sqrt{3}\\
	 \sqrt{3} & 3 & -2 & -2 & 1 & -2\sqrt{3} & 0 & - \sqrt{3}\\
	3 &  \sqrt{3} & -\sqrt{3} & \sqrt{3} & 0 & 3 & 3 & 0\\
	0 &  -2\sqrt{3} & -2\sqrt{3} & 0 &  -\sqrt{3} & 0 & 0 & -3\\
	-3 &  \sqrt{3} & 0 & 2\sqrt{3} &  \sqrt{3} & 0 & 0 & -3
\end{matrix}\right).
\end{split}
\end{align}
The octet is a real representation.

\begin{flalign}\label{eq:tensor23b}
&\text{{\textit {\textbf {iii}}})\,\,\,}\boldsymbol{2}\otimes\boldsymbol{\tb}=\boldsymbol{6}&
\end{flalign}
We define the basis for the doublet representation in such a way that $\boldsymbol{6}$ is simply the Kronecker product of $\boldsymbol{2}$ and $\boldsymbol{\tb}$, i.e.
\begin{equation}\label{eq:tensor2exp}
\boldsymbol{6}\equiv\left(\begin{matrix}a_1 \B_1 \\
	a_1 \B_2\\
	a_1 \B_3\\
	a_2 \B_1\\
	a_2 \B_2\\
        a_2 \B_3
\end{matrix}\right)
\end{equation}
where $(a_1, a_2)^T$ and $(\B_1, \B_2, \B_3)^T$ represent $\boldsymbol{2}$ and $\boldsymbol{\tb}$ respectively. In such a basis, the generator matrices for the doublet are
\begin{align}\label{eq:gen2}
\begin{split}
&C \equiv
\left(\begin{matrix}1 & 0\\
	0 & 1
\end{matrix}\right), \quad  \quad  \quad \quad \, \, \, E \equiv
\left(\begin{matrix}1 & 0\\
	0 & 1
\end{matrix}\right),\\
&V \equiv
\frac{i}{\sqrt{3}}\left(\begin{matrix}1 & \sqrt{2}\\
	 \sqrt{2} & -1
\end{matrix}\right), \quad X \equiv
\frac{i}{\sqrt{3}}\left(\begin{matrix}1 & \sqrt{2}\ob\\
	 \sqrt{2}\om & -1
\end{matrix}\right).\\
\end{split}
\end{align}

\begin{flalign}\label{eq:tensor22}
&\text{{\textit {\textbf {iv}}})\,\,\,}\boldsymbol{2}\otimes\boldsymbol{2}=\boldsymbol{1}\oplus \boldsymbol{1^{(0,1)}}\oplus\boldsymbol{1^{(1,0)}}\oplus\boldsymbol{1^{(1,1)}}&
\end{flalign}
The singlets $\boldsymbol{1^{(p,q)}}$ transform as
\begin{equation}\label{eq:tensor22exp}
C \equiv 1, \quad E \equiv 1, \quad V \equiv (-1)^p, \quad X \equiv (-1)^q. 
\end{equation}
In terms of the tensor product expansion, Eq.~(\ref{eq:tensor22}), these singlets are given by
\begin{align}\label{eq:tensor22cg1}
\begin{split}
&\boldsymbol{1} \equiv a^T u \, b,\\
&\boldsymbol{1^{(0,1)}} \equiv a^T u_1 \, b, \,\,\, \boldsymbol{1^{(1,0)}}\equiv a^T u_\om \, b, \,\,\, \boldsymbol{1^{(1,1)}} \equiv a^T u_{\ob} \, b\\
\end{split}
\end{align}
where $a$ and $b$ represent the doublets in Eq.~(\ref{eq:tensor22}) and $u$, $u_1$, $u_{\om}$ and $u_{\ob}$ are unitary matrices,
\begin{align}\label{eq:tensor22cg2}
\begin{split}
&u=i \sigma_2\\
&u_1 = u V, \quad u_{\om}= u X, \quad u_{\ob}= -u \bar{X},\\
\end{split}
\end{align}
with $\sigma_2$ being the second Pauli matrix and $V$, $X$ being the generators of the doublet representation, Eq~(\ref{eq:gen2}).
\begin{comment}
\begin{align}\label{eq:tensor22cg}
\begin{split}
&\boldsymbol{1} \equiv \frac{1}{\sqrt{2}}(a_1 b_2 - a_2 b_1),\\
&\boldsymbol{1^{(0,1)}} \equiv \frac{1}{\sqrt{3}} a_1 b_1 -\frac{1}{\sqrt{6}}a_2 b_1-\frac{1}{\sqrt{6}}a_1 b_2-\frac{1}{\sqrt{3}}a_2 b_2,\\
&\boldsymbol{1^{(1,0)}} \equiv \frac{\om}{\sqrt{3}} a_1 b_1 -\frac{1}{\sqrt{6}}a_2 b_1-\frac{1}{\sqrt{6}}a_1 b_2-\frac{\ob}{\sqrt{3}}a_2 b_2,\\
&\boldsymbol{1^{(1,1)}} \equiv \frac{\ob}{\sqrt{3}} a_1 b_1 -\frac{1}{\sqrt{6}}a_2 b_1-\frac{1}{\sqrt{6}}a_1 b_2-\frac{\om}{\sqrt{3}}a_2 b_2\\
\end{split}
\end{align}
where $(a_1, a_2)^T$ and $(b_1, b_2)^T$ represent the doublets appearing in the LHS of Eq.~(\ref{eq:tensor22})
\end{comment}

\begin{flalign}\label{eq:tensor63}
&\text{{\textit {\textbf {v}}})\,\,\,}\boldsymbol{6}\otimes\boldsymbol{3}=\boldsymbol{2}\oplus\boldsymbol{8}\oplus\boldsymbol{8}&
\end{flalign}
The C-G coefficients for the above tensor product expansion are given by
\begin{flalign}\label{eq:tensor63exp1}
&\boldsymbol{2}\equiv\left(\begin{matrix}
\frac{1}{\sqrt{3}}a_1 b_1 +\frac{1}{\sqrt{3}}a_2 b_2+\frac{1}{\sqrt{3}}a_3 b_3\\
\frac{1}{\sqrt{3}}a_4 b_1 +\frac{1}{\sqrt{3}}a_5 b_2+\frac{1}{\sqrt{3}}a_6 b_3\\
\end{matrix}\right),&
\end{flalign}
\begin{flalign}\label{eq:tensor63exp2}
&\boldsymbol{8}\equiv\left(\begin{matrix}
-\frac{1}{\sqrt{2}}a_1 b_1+\frac{1}{\sqrt{2}}a_3 b_3\\
\frac{1}{\sqrt{6}}a_1 b_1-\frac{\sqrt{2}}{\sqrt{3}}a_2 b_2+\frac{1}{\sqrt{6}}a_3 b_3\\
\frac{1}{\sqrt{6}}a_2 b_3-\frac{1}{\sqrt{6}}a_3 b_2+\frac{1}{\sqrt{3}}a_4 b_2-\frac{1}{\sqrt{3}}a_4 b_3\\
\frac{1}{\sqrt{6}}a_3 b_1-\frac{1}{\sqrt{6}}a_1 b_3+\frac{1}{\sqrt{3}}a_5 b_3-\frac{1}{\sqrt{3}}a_5 b_1\\
\frac{1}{\sqrt{6}}a_1 b_2-\frac{1}{\sqrt{6}}a_2 b_1+\frac{1}{\sqrt{3}}a_6 b_1-\frac{1}{\sqrt{3}}a_6 b_2\\
-\frac{i}{\sqrt{6}}a_2 b_3-\frac{i}{\sqrt{6}}a_3 b_2-\frac{i}{\sqrt{3}}a_4 b_2-\frac{i}{\sqrt{3}}a_4 b_3\\
-\frac{i}{\sqrt{6}}a_3 b_1-\frac{i}{\sqrt{6}}a_1 b_3-\frac{i}{\sqrt{3}}a_5 b_3-\frac{i}{\sqrt{3}}a_5 b_1\\
-\frac{i}{\sqrt{6}}a_1 b_2-\frac{i}{\sqrt{6}}a_2 b_1-\frac{i}{\sqrt{3}}a_6 b_1-\frac{i}{\sqrt{3}}a_6 b_2\\
\end{matrix}\right),&
\end{flalign}
\begin{flalign}\label{eq:tensor63exp3}
&\boldsymbol{8}\equiv\left(\begin{matrix}
-\frac{1}{\sqrt{2}}a_4 b_1+\frac{1}{\sqrt{2}}a_6 b_3\\
\frac{1}{\sqrt{6}}a_4 b_1-\frac{\sqrt{2}}{\sqrt{3}}a_5 b_2+\frac{1}{\sqrt{6}}a_6 b_3\\
-\frac{1}{\sqrt{6}}a_5 b_3+\frac{1}{\sqrt{6}}a_6 b_2-\frac{1}{\sqrt{3}}a_2 b_1+\frac{1}{\sqrt{3}}a_3 b_1\\
-\frac{1}{\sqrt{6}}a_6 b_1+\frac{1}{\sqrt{6}}a_4 b_3-\frac{1}{\sqrt{3}}a_3 b_2+\frac{1}{\sqrt{3}}a_1 b_2\\
-\frac{1}{\sqrt{6}}a_4 b_2+\frac{1}{\sqrt{6}}a_5 b_1-\frac{1}{\sqrt{3}}a_1 b_3+\frac{1}{\sqrt{3}}a_2 b_3\\
\frac{i}{\sqrt{6}}a_5 b_3+\frac{i}{\sqrt{6}}a_6 b_2-\frac{i}{\sqrt{3}}a_2 b_1-\frac{i}{\sqrt{3}}a_3 b_1\\
\frac{i}{\sqrt{6}}a_6 b_1+\frac{i}{\sqrt{6}}a_4 b_3-\frac{i}{\sqrt{3}}a_3 b_2-\frac{i}{\sqrt{3}}a_1 b_2\\
\frac{i}{\sqrt{6}}a_4 b_2+\frac{i}{\sqrt{6}}a_5 b_1-\frac{i}{\sqrt{3}}a_1 b_3-\frac{i}{\sqrt{3}}a_2 b_3\\
\end{matrix}\right),&
\end{flalign}
where $(a_1, a_2, a_3, a_4, a_5, a_6)^T$ and $(b_1, b_2, b_3)^T$ represent the sextet and the triplet appearing in the LHS of Eq.~(\ref{eq:tensor63}).

\begin{flalign}\label{eq:tensor63b}
&\text{{\textit {\textbf {vi}}})\,\,\,}\boldsymbol{6}\otimes\boldsymbol{\tb}=\boldsymbol{3}\oplus\boldsymbol{\xb}\oplus\boldsymbol{3^{(0,1)}}\oplus\boldsymbol{3^{(1,0)}}\oplus\boldsymbol{3^{(1,1)}}&
\end{flalign}
The representations $\boldsymbol{3^{(0,1)}}$, $\boldsymbol{3^{(1,0)}}$ and  $\boldsymbol{3^{(1,1)}}$ are simply the product of the triplet $\boldsymbol{3}$ and the singlets $\boldsymbol{1^{(0,1)}}$, $\boldsymbol{1^{(1,0)}}$ and $\boldsymbol{1^{(1,1)}}$ respectively, 
\begin{equation}\label{eq:tensor66hes}
\boldsymbol{3^{(p,q)}}=\boldsymbol{3}\,\boldsymbol{1^{(p,q)}}.
\end{equation}
The C-G coefficients for the tensor product expansion, Eq.~(\ref{eq:tensor63b}), are given by
\begin{flalign}\label{eq:tensor63bexp1}
&\boldsymbol{3}\equiv\left(\begin{matrix}
\frac{1}{\sqrt{2}}a_1 \B_1 +\frac{1}{2}a_5 \B_3+\frac{1}{2}a_6 \B_2\\
\frac{1}{\sqrt{2}}a_2 \B_2 +\frac{1}{2}a_6 \B_1+\frac{1}{2}a_4 \B_3\\
\frac{1}{\sqrt{2}}a_3 \B_3 +\frac{1}{2}a_4 \B_2+\frac{1}{2}a_5 \B_1\\
\end{matrix}\right),&
\end{flalign}
\begin{flalign}\label{eq:tensor63bexp2}
&\boldsymbol{\xb}\equiv\left(\begin{matrix}
-\frac{1}{\sqrt{2}}a_5 \B_3 +\frac{1}{\sqrt{2}}a_6 \B_2\\
-\frac{1}{\sqrt{2}}a_6 \B_1 +\frac{1}{\sqrt{2}}a_4 \B_3\\
-\frac{1}{\sqrt{2}}a_4 \B_2 +\frac{1}{\sqrt{2}}a_5 \B_1\\
\frac{1}{\sqrt{2}}a_2 \B_3 -\frac{1}{\sqrt{2}}a_3 \B_2\\
\frac{1}{\sqrt{2}}a_3 \B_1 -\frac{1}{\sqrt{2}}a_1 \B_3\\
\frac{1}{\sqrt{2}}a_1 \B_2 -\frac{1}{\sqrt{2}}a_2 \B_1\\
\end{matrix}\right),&
\end{flalign}
\begin{flalign}\label{eq:tensor63bexp3}
&\boldsymbol{3^{(0,1)}}\equiv\left(\begin{matrix}
 \frac{1}{\sqrt{6}}a_1 \B_1+\frac{1}{\sqrt{6}}a_{\{2} \B_{3\}}+\frac{1}{\sqrt{3}}a_4 \B_1-\frac{1}{2\sqrt{3}}(a_5 \B_3+a_6 \B_2)\\
\frac{1}{\sqrt{6}}a_2 \B_2+\frac{1}{\sqrt{6}}a_{\{3} \B_{1\}}+\frac{1}{\sqrt{3}}a_5 \B_2-\frac{1}{2\sqrt{3}}(a_6 \B_1+a_4 \B_3)\\
\frac{1}{\sqrt{6}}a_3 \B_3+\frac{1}{\sqrt{6}}a_{\{1} \B_{2\}}+\frac{1}{\sqrt{3}}a_6 \B_3-\frac{1}{2\sqrt{3}}(a_4 \B_2+a_5 \B_1)\\
\end{matrix}\right),&
\end{flalign}
\begin{flalign}\label{eq:tensor63bexp4}
&\boldsymbol{3^{(1,0)}}\equiv\left(\begin{matrix}
 \frac{1}{\sqrt{6}}a_1 \B_1+\frac{\om}{\sqrt{6}}a_{\{2} \B_{3\}}+\frac{\ob}{\sqrt{3}}a_4 \B_1-\frac{1}{2\sqrt{3}}(a_5 \B_3+a_6 \B_2)\\
\frac{1}{\sqrt{6}}a_2 \B_2+\frac{\om}{\sqrt{6}}a_{\{3} \B_{1\}}+\frac{\ob}{\sqrt{3}}a_5 \B_2-\frac{1}{2\sqrt{3}}(a_6 \B_1+a_4 \B_3)\\
\frac{1}{\sqrt{6}}a_3 \B_3+\frac{\om}{\sqrt{6}}a_{\{1} \B_{2\}}+\frac{\ob}{\sqrt{3}}a_6 \B_3-\frac{1}{2\sqrt{3}}(a_4 \B_2+a_5 \B_1)\\
\end{matrix}\right),&
\end{flalign}
\begin{flalign}\label{eq:tensor63bexp5}
&\boldsymbol{3^{(1,1)}}\equiv\left(\begin{matrix}
 \frac{1}{\sqrt{6}}a_1 \B_1+\frac{\ob}{\sqrt{6}}a_{\{2} \B_{3\}}+\frac{\om}{\sqrt{3}}a_4 \B_1-\frac{1}{2\sqrt{3}}(a_5 \B_3+a_6 \B_2)\\
\frac{1}{\sqrt{6}}a_2 \B_2+\frac{\ob}{\sqrt{6}}a_{\{3} \B_{1\}}+\frac{\om}{\sqrt{3}}a_5 \B_2-\frac{1}{2\sqrt{3}}(a_6 \B_1+a_4 \B_3)\\
\frac{1}{\sqrt{6}}a_3 \B_3+\frac{\ob}{\sqrt{6}}a_{\{1} \B_{2\}}+\frac{\om}{\sqrt{3}}a_6 \B_3-\frac{1}{2\sqrt{3}}(a_4 \B_2+a_5 \B_1)\\
\end{matrix}\right),&
\end{flalign}
where $(a_1, a_2, a_3, a_4, a_5, a_6)^T$ and $(\B_1, \B_2, \B_3)^T$ represent the sextet and the conjugate triplet appearing in the LHS of Eq.~(\ref{eq:tensor63b}). In Eqs.~(\ref{eq:tensor63bexp3}-\ref{eq:tensor63bexp5}) we have used the curly bracket to denote the symmetric sum, i.e.~$a_{\{i} \B_{j\}}=a_i \B_j + a_j \B_i$. 

\begin{flalign}\label{eq:tensor66}
&\text{{\textit {\textbf {vii}}})\,\,\,}\boldsymbol{6}\otimes\boldsymbol{6}=\underbrace{\boldsymbol{\xb}\oplus\boldsymbol{\xb}\oplus\boldsymbol{\xb}\oplus\boldsymbol{3}}_\text{symm}\oplus\underbrace{\boldsymbol{\xb}\oplus\boldsymbol{3^{(0,1)}}\oplus\boldsymbol{3^{(1,0)}}\oplus\boldsymbol{3^{(1,1)}}}_\text{antisymm}&
\end{flalign}
Here the sextet, $\boldsymbol{\xb}$, appears more than once in the symmetric part. So there is no unique way of decomposing the product space into the sum of the irreducible sextets, i.e.~the C-G coefficients are not uniquely defined. To solve this problem, we utilise the group $\Sigma(216\times3)$ which has $\Sigma(72\times3)$ as one of its subgroups. $\Sigma(216\times3)$ has three distinct types of sextets~\cite{Sigma1}, $\boldsymbol{6^0}$, $\boldsymbol{6^1}$, $\boldsymbol{6^2}$. The sextet of $\Sigma(72\times3)$ can be embedded in any of these three sextets of $\Sigma(216\times3)$. The tensor product expansion for two $\boldsymbol{6^0}$s of $\Sigma(216\times3)$ is given by 
\begin{equation}\label{eq:tensor66hes}
\boldsymbol{6^0}\otimes\boldsymbol{6^0}=\underbrace{\boldsymbol{\xb^0}\oplus\boldsymbol{\xb^1}\oplus\boldsymbol{\xb^2}\oplus\boldsymbol{3^0}}_\text{symm}\oplus\underbrace{\boldsymbol{\xb^0}\oplus\boldsymbol{9}}_\text{antisymm}.
\end{equation}
In Eq.~(\ref{eq:tensor66hes}), the decomposition of the symmetric part into the irreducible sextets is unique. Hence we embed the irreps of $\Sigma(72\times3)$ in the irreps of $\Sigma(216\times3)$,
{\small
\begin{equation}\label{eq:orig}
\begin{tikzpicture}[>=stealth,baseline=(current bounding box.center),anchor=base,inner sep=0pt]
\matrix (foil) [matrix of math nodes,nodes={minimum height=1em}] {
\Sigma(216\times3)&\,:\,& \boldsymbol{6^0}&\otimes&\boldsymbol{6^0}&=&\boldsymbol{\bar{6^0}}&\oplus&\boldsymbol{\bar{6^1}}&\oplus&\boldsymbol{\bar{6^2}}&\oplus&\boldsymbol{3^0}&\oplus&\boldsymbol{\bar{6^0}}&\oplus&\boldsymbol{9}\\
\,\\
\Sigma(72\times3)&\,:\,& \boldsymbol{6}&\otimes&\boldsymbol{6}&=&\boldsymbol{\xb}&\oplus&\boldsymbol{\xb}&\oplus&\boldsymbol{\xb}&\oplus&\boldsymbol{3}&\oplus&\boldsymbol{\xb}&\oplus&\boldsymbol{3^{(0,1)}}&\oplus&\boldsymbol{3^{(1,0)}}&\oplus&\boldsymbol{3^{(1,1)}},\\};
\path[->] ($(foil-1-1.south)$)  edge[]     ($(foil-3-1.north)$);
\path[->] ($(foil-1-3.south)$)  edge[]     ($(foil-3-3.north)$);
\path[->] ($(foil-1-5.south)$)   edge[]    ($(foil-3-5.north)$);
\path[->] ($(foil-1-7.south)$)  edge[]     ($(foil-3-7.north)$);
\path[->] ($(foil-1-9.south)$)   edge[]    ($(foil-3-9.north)$);
\path[->] ($(foil-1-11.south)$)  edge[]     ($(foil-3-11.north)$);
\path[->] ($(foil-1-13.south)$)   edge[]    ($(foil-3-13.north)$);
\path[->] ($(foil-1-15.south)$)  edge[]     ($(foil-3-15.north)$);
\path[->] ($(foil-1-17.south)$)   edge[]    ($(foil-3-17.north)$);
\path[->] ($(foil-1-17.south)$)  edge[]     ($(foil-3-19.north)$);
\path[->] ($(foil-1-17.south)$)   edge[]    ($(foil-3-21.north)$);
\end{tikzpicture}
\end{equation}
}to obtain a unique decomposition for the case of $\Sigma(72\times3)$ as well. Thus the C-G coefficients for Eq.~(\ref{eq:tensor66}) are given by

\begin{flalign}\label{eq:tensor66exp1}
&\boldsymbol{\xb}\equiv\left(\begin{matrix}\frac{1}{\sqrt{3}}a_{\{2}b_{3\}}-\frac{1}{\sqrt{3}}a_4 b_4\\
\frac{1}{\sqrt{3}}a_{\{3}b_{1\}}-\frac{1}{\sqrt{3}}a_5 b_5\\
\frac{1}{\sqrt{3}}a_{\{1}b_{2\}}-\frac{1}{\sqrt{3}}a_6 b_6\\
\frac{1}{\sqrt{6}}a_{\{5}b_{6\}}-\frac{1}{\sqrt{3}}a_{\{1}b_{4\}}\\
\frac{1}{\sqrt{6}}a_{\{6}b_{4\}}-\frac{1}{\sqrt{3}}a_{\{2}b_{5\}}\\
\frac{1}{\sqrt{6}}a_{\{4}b_{5\}}-\frac{1}{\sqrt{3}}a_{\{3}b_{6\}}\\
\end{matrix}\right),&
\end{flalign}
\begin{flalign}\label{eq:tensor66exp2}
&\boldsymbol{\xb}\equiv\left(\begin{matrix}-\frac{1}{\sqrt{3}}a_1 b_1+\frac{1}{\sqrt{6}}a_{\{2}b_{6\}}+\frac{1}{\sqrt{6}}a_{\{3}b_{5\}}\\
-\frac{1}{\sqrt{3}}a_2 b_2+\frac{1}{\sqrt{6}}a_{\{3}b_{4\}}+\frac{1}{\sqrt{6}}a_{\{1}b_{6\}}\\
-\frac{1}{\sqrt{3}}a_3 b_3+\frac{1}{\sqrt{6}}a_{\{1}b_{5\}}+\frac{1}{\sqrt{6}}a_{\{2}b_{4\}}\\
\frac{\sqrt{2}}{\sqrt{3}}a_4 b_4+\frac{1}{\sqrt{6}}a_{\{2}b_{3\}}\\
\frac{\sqrt{2}}{\sqrt{3}}a_5 b_5+\frac{1}{\sqrt{6}}a_{\{3}b_{1\}}\\
\frac{\sqrt{2}}{\sqrt{3}}a_6 b_6+\frac{1}{\sqrt{6}}a_{\{1}b_{2\}}\\
\end{matrix}\right),&
\end{flalign}
\begin{flalign}\label{eq:tensor66exp3}
&\boldsymbol{\xb}\equiv\left(\begin{matrix}\frac{1}{\sqrt{6}}a_{\{1}b_{4\}}+\frac{1}{\sqrt{3}}a_{\{5}b_{6\}}\\
\frac{1}{\sqrt{6}}a_{\{2}b_{5\}}+\frac{1}{\sqrt{3}}a_{\{6}b_{4\}}\\
\frac{1}{\sqrt{6}}a_{\{3}b_{6\}}+\frac{1}{\sqrt{3}}a_{\{4}b_{5\}}\\
-\frac{\sqrt{2}}{\sqrt{3}}a_1 b_1-\frac{1}{2\sqrt{3}}a_{\{2}b_{6\}}-\frac{1}{2\sqrt{3}}a_{\{3}b_{5\}}\\
-\frac{\sqrt{2}}{\sqrt{3}}a_2 b_2-\frac{1}{2\sqrt{3}}a_{\{3}b_{4\}}-\frac{1}{2\sqrt{3}}a_{\{1}b_{6\}}\\
-\frac{\sqrt{2}}{\sqrt{3}}a_3 b_3-\frac{1}{2\sqrt{3}}a_{\{1}b_{5\}}-\frac{1}{2\sqrt{3}}a_{\{2}b_{4\}}\\
\end{matrix}\right),&
\end{flalign}
\begin{flalign}\label{eq:tensor66exp4}
&\boldsymbol{3}\equiv\left(\begin{matrix}
\frac{1}{2}a_{\{2}b_{6\}}-\frac{1}{2}a_{\{3}b_{5\}}\\
\frac{1}{2}a_{\{3}b_{4\}}-\frac{1}{2}a_{\{1}b_{6\}}\\
\frac{1}{2}a_{\{1}b_{5\}}-\frac{1}{2}a_{\{2}b_{4\}}\\
\end{matrix}\right),&
\end{flalign}
\begin{flalign}\label{eq:tensor66exp5}
&\boldsymbol{\xb}\equiv\left(\begin{matrix}
\frac{1}{\sqrt{2}}a_{[1}b_{4]}\\
\frac{1}{\sqrt{2}}a_{[2}b_{5]}\\
\frac{1}{\sqrt{2}}a_{[3}b_{6]}\\
\frac{1}{2}a_{[2}b_{6]}+\frac{1}{2}a_{[3}b_{5]}\\
\frac{1}{2}a_{[3}b_{4]}+\frac{1}{2}a_{[1}b_{6]}\\
\frac{1}{2}a_{[1}b_{5]}+\frac{1}{2}a_{[2}b_{4]}\\
\end{matrix}\right),&
\end{flalign}
\begin{flalign}\label{eq:tensor66exp6}
&\boldsymbol{3^{(0,1)}}\equiv\left(\begin{matrix}
-\frac{1}{\sqrt{6}}a_{[2}b_{3]}+\frac{1}{2\sqrt{3}}a_{[2}b_{6]}-\frac{1}{2\sqrt{3}}a_{[3}b_{5]}+\frac{1}{\sqrt{6}}a_{[5}b_{6]}\\
-\frac{1}{\sqrt{6}}a_{[3}b_{1]}+\frac{1}{2\sqrt{3}}a_{[3}b_{4]}-\frac{1}{2\sqrt{3}}a_{[1}b_{6]}+\frac{1}{\sqrt{6}}a_{[6}b_{4]}\\
-\frac{1}{\sqrt{6}}a_{[1}b_{2]}+\frac{1}{2\sqrt{3}}a_{[1}b_{5]}-\frac{1}{2\sqrt{3}}a_{[2}b_{4]}+\frac{1}{\sqrt{6}}a_{[4}b_{5]}\\
\end{matrix}\right),&
\end{flalign}
\begin{flalign}\label{eq:tensor66exp7}
&\boldsymbol{3^{(1,0)}}\equiv\left(\begin{matrix}
-\frac{\om}{\sqrt{6}}a_{[2}b_{3]}+\frac{1}{2\sqrt{3}}a_{[2}b_{6]}-\frac{1}{2\sqrt{3}}a_{[3}b_{5]}+\frac{\ob}{\sqrt{6}}a_{[5}b_{6]}\\
-\frac{\om}{\sqrt{6}}a_{[3}b_{1]}+\frac{1}{2\sqrt{3}}a_{[3}b_{4]}-\frac{1}{2\sqrt{3}}a_{[1}b_{6]}+\frac{\ob}{\sqrt{6}}a_{[6}b_{4]}\\
-\frac{\om}{\sqrt{6}}a_{[1}b_{2]}+\frac{1}{2\sqrt{3}}a_{[1}b_{5]}-\frac{1}{2\sqrt{3}}a_{[2}b_{4]}+\frac{\ob}{\sqrt{6}}a_{[4}b_{5]}\\
\end{matrix}\right),&
\end{flalign}
\begin{flalign}\label{eq:tensor66exp8}
&\boldsymbol{3^{(1,1)}}\equiv\left(\begin{matrix}
-\frac{\ob}{\sqrt{6}}a_{[2}b_{3]}+\frac{1}{2\sqrt{3}}a_{[2}b_{6]}-\frac{1}{2\sqrt{3}}a_{[3}b_{5]}+\frac{\om}{\sqrt{6}}a_{[5}b_{6]}\\
-\frac{\ob}{\sqrt{6}}a_{[3}b_{1]}+\frac{1}{2\sqrt{3}}a_{[3}b_{4]}-\frac{1}{2\sqrt{3}}a_{[1}b_{6]}+\frac{\om}{\sqrt{6}}a_{[6}b_{4]}\\
-\frac{\ob}{\sqrt{6}}a_{[1}b_{2]}+\frac{1}{2\sqrt{3}}a_{[1}b_{5]}-\frac{1}{2\sqrt{3}}a_{[2}b_{4]}+\frac{\om}{\sqrt{6}}a_{[4}b_{5]}\\
\end{matrix}\right)&
\end{flalign}
where $(a_1, a_2, a_3, a_4, a_5, a_6)^T$ and $(b_1, b_2, b_3, b_4, b_5, b_6)^T$ represent the sextets appearing in the LHS of Eq.~(\ref{eq:tensor66}). In Eqs.~(\ref{eq:tensor66exp1}-\ref{eq:tensor66exp8}) we have used the curly bracket and the square bracket to denote the symmetric sum and the antisymmetric sum respectively, i.e.~$a_{\{i} b_{j\}}=a_i b_j + a_j b_i$ and $a_{[i} b_{j]}=a_i b_j - a_j b_i$ .

\begin{flalign}\label{eq:tensor66b}
&\text{{\textit {\textbf {viii}}})\,\,\,}\boldsymbol{6}\otimes\boldsymbol{\xb}=\boldsymbol{1}\oplus\boldsymbol{8}\oplus\boldsymbol{1^{(0,1)}}\oplus\boldsymbol{1^{(1,0)}}\oplus\boldsymbol{1^{(1,1)}}\oplus\boldsymbol{8}\oplus\boldsymbol{8}\oplus\boldsymbol{8}&
\end{flalign}
We are not listing the C-G coefficients for the above expansion, since they are not used in our model.

%----------------------------------------------------------------------------------------------------------------------------------
\section*{Appendix B: \quad Hierarchical Structure \,of\, the Charged-Lepton Mass Matrix}

The triplet flavons, $\phi_\alpha$ and $\phi_\beta$, transform as $\boldsymbol{3}\times-i$ and $\boldsymbol{3}\times i$ under $\Sigma(72\times3)\times C_4$, Table~\ref{tab:flavourcontent}. $L^\dagger\tau_R$, $L^\dagger\mu_R$, $L^\dagger e_R$ transform as $\boldsymbol{3}\times i$, $\boldsymbol{3}\times 1$, $\boldsymbol{3}\times -1$ respectively. Therefore, the flavons and their tensor products which transform as $\boldsymbol{\tb}$ under $\Sigma(72\times3)$ and $-i$, $1$, $-1$ under $C_4$ couple with $L^\dagger\tau_R$, $L^\dagger\mu_R$, $L^\dagger e_R$ respectively. $C_4$ is responsible for restricting the allowed couplings and produces the hierarchical structure of the mass matrix. In Section~3, we showed that $\bar{\phi}_\beta$ and $\bar{A}_{\beta\alpha}$ couple to the $\tau$ and $\mu$ sectors. After symmetry breaking, the flavons attain the VEVs $\langle\phi_\alpha\rangle =V^\dagger(1,0,0)^Tm$ and $\langle\phi_\beta\rangle =V^\dagger(0,0,1)^Tm$, Eq.~(\ref{eq:leptvev}). The resulting tau and muon masses are of the order of $\epsilon$ and $\epsilon^2$ respectively. The term ${\mathcal H}. {\mathcal T.}$ in Eq.~(\ref{eq:mclept}) contains all the higher order products of the flavons transforming as $\boldsymbol{\tb}$ and $-i$, $1$, $-1$ coupling to $\tau$, $\mu$, $e$ sectors. In this Appendix, we analyse the cubic and the quartic products which give rise ${\mathcal O}(\epsilon^3)$ and ${\mathcal O}(\epsilon^4)$ mass matrix elements respectively in Eq.~(\ref{eq:leptcontrib}). We neglect the products beyond quartic order.

\subsection*{Cubic Products}

\begin{flalign}
\text{{\textit {\textbf {i}}})\,\,}\boldsymbol{3}\otimes\boldsymbol{3}\otimes\boldsymbol{3}
&=(\boldsymbol{6}\oplus\boldsymbol{\tb})\otimes\boldsymbol{3}&\notag\\
&=\boldsymbol{2}\oplus\boldsymbol{8}\oplus\boldsymbol{8}\oplus\boldsymbol{1}
\oplus\boldsymbol{8}&\label{eq:cubic1}\\
\text{{\textit {\textbf {ii}}})\,\,\,}\boldsymbol{\tb}\otimes\boldsymbol{\tb}\otimes\boldsymbol{\tb}
&=(\boldsymbol{\xb}\oplus\boldsymbol{3})\otimes\boldsymbol{\tb}&\notag\\
&=\boldsymbol{2}\oplus\boldsymbol{8}\oplus\boldsymbol{8}\oplus\boldsymbol{1}
\oplus\boldsymbol{8}&\label{eq:cubic2}\\
\text{{\textit {\textbf {iii}}})\,\,}\boldsymbol{3}\otimes\boldsymbol{3}\otimes\boldsymbol{\tb}
&=(\boldsymbol{6}\oplus\boldsymbol{\tb})\otimes\boldsymbol{\tb}&\notag\\
&=\boldsymbol{3}\oplus\boldsymbol{\xb}\oplus\boldsymbol{3^{(0,1)}}\oplus\boldsymbol{3^{(1,0)}}\oplus\boldsymbol{3^{(1,1)}}
\oplus\boldsymbol{\xb}\oplus\boldsymbol{3}&\label{eq:cubic3}
\end{flalign}
The above expansions, Eqs.~(\ref{eq:cubic1}-\ref{eq:cubic3}), do not contribute to any coupling, since $\boldsymbol{\bar{3}}$ does not appear in their RHS.

\begin{flalign}\label{eq:cubic4}
\text{{\textit {\textbf {iv}}})\,\,}\boldsymbol{\tb}\otimes\boldsymbol{\tb}\otimes\boldsymbol{3}
&=(\boldsymbol{\xb}\oplus\boldsymbol{3})\otimes\boldsymbol{3}&\notag\\
&=\boldsymbol{\tb}\oplus\boldsymbol{6}\oplus\boldsymbol{\ab}\oplus\boldsymbol{\bb}\oplus\boldsymbol{\cb}
\oplus\boldsymbol{6}\oplus\boldsymbol{\tb}&
\end{flalign}
In the above expansion, $\boldsymbol{\tb}$ appears twice in the RHS. In terms of the components of the triplets, these $\boldsymbol{\tb}$s are given by
\begin{align}
\begin{split}\label{eq:cubic31}
1^\text{st}\,\boldsymbol{\tb}=&\frac{1}{2\sqrt{2}}\left(\A_1 (2\B_1 c_1+\B_2 c_2+\B_3 c_3)+\B_1(\A_2 c_2+\A_3 c_3),\right. \\
&\left. \quad\quad\,\, \A_2 (2\B_2 c_2+\B_3 c_3+\B_1 c_1)+\B_2(\A_3 c_3+\A_1 c_1),\right. \\
&\left. \quad\quad\,\, \A_3 (2\B_3 c_3+\B_1 c_1+\B_2 c_2)+\B_3( \A_1 c_1+\A_2 c_2)\right)^T,
\end{split}\\
\begin{split}\label{eq:cubic32}
2^\text{nd}\,\boldsymbol{\tb}=&\frac{1}{2}\left(\B_1 (\A_2 c_2+\A_3 c_3)-\A_1 (\B_2 c_2+\B_3 c_3),\right. \\
&\left. \quad\,\, \B_2 (\A_3 c_3+\A_1 c_1)-\A_2 (\B_3 c_3+\B_1 c_1) ,\right. \\
&\left. \quad\,\, \B_3 (\A_1 c_1+\A_2 c_2)-\A_3 (\B_1 c_1+\B_2 c_2)\right)^T
\end{split}
\end{align}
where $(\A_1, \A_2, \A_3)^T$, $(\B_1, \B_2, \B_3)^T$ and $(c_1, c_2, c_3)^T$ are the $\boldsymbol{\tb}$, $\boldsymbol{\tb}$ and $\boldsymbol{3}$ appearing in the LHS of Eq.~(\ref{eq:cubic4}). The product $\boldsymbol{\tb}\otimes\boldsymbol{\tb}\otimes\boldsymbol{3}$ can be obtained in terms of the flavons $\phi_\alpha$ and $\phi_\beta$ in several different ways. These are listed in Table~\ref{tab:cubic}. For each combination of flavons, we provide the corresponding $C_4$ representation. Under $\langle\phi_\alpha\rangle\propto(1,0,0)$ and $\langle\phi_\beta\rangle\propto(0,0,1)$\footnote{For the sake of brevity, in this Appendix we omit $V^\dagger$, $m$ and the transposition when referring to the VEVs, i.e.~$(1,0,0)\equiv V^\dagger(1,0,0)^Tm$}, we calculate the vacuum alignments of the cubic $\boldsymbol{\bar{3}}$s given in Eq.~(\ref{eq:cubic31}) and Eq.~(\ref{eq:cubic32}). These are listed in the last two columns of the table.

{\renewcommand{\arraystretch}{1.5}
\begin{table}[]
\begin{center}
\begin{tabular}{|c|c|c c|}
\hline
 &$C_4$	&$1^\text{st}\,\boldsymbol{\tb}$	&$2^\text{nd}\,\boldsymbol{\tb}$\\
\hline
$\bar{\phi}_\alpha \bar{\phi}_\alpha \phi_\alpha$
&$i$&$(\frac{1}{\sqrt{2}},0,0)$&$(0,0,0)$\\
$\bar{\phi}_\alpha \bar{\phi}_\alpha \phi_\beta$
&$-i$&$(0,0,0)$&$(0,0,0)$\\
$\bar{\phi}_\alpha \bar{\phi}_\beta \phi_\alpha$
&$-i$&$(0,0,\frac{1}{2\sqrt{2}})$&$(0,0,\frac{1}{2})$\\
$\bar{\phi}_\alpha \bar{\phi}_\beta \phi_\beta$
&$i$&$(\frac{1}{2\sqrt{2}},0,0)$&$(-\frac{1}{2},0,0)$\\
$\bar{\phi}_\beta \bar{\phi}_\beta \phi_\alpha$
&$i$&$(0,0,0)$&$(0,0,0)$\\
$\bar{\phi}_\beta \bar{\phi}_\beta \phi_\beta$
&$-i$&$(0,0,\frac{1}{\sqrt{2}})$&$(0,0,0)$\\
\hline
\end{tabular}
\end{center}
\caption{Cubic products of $\phi_\alpha$ and $\phi_\beta$ of the form $\boldsymbol{\tb}\otimes\boldsymbol{\tb}\otimes\boldsymbol{3}$ leading to $\boldsymbol{\bar{3}}$s.}
\label{tab:cubic}
\end{table}

The products transforming as $i$ under $C_4$ can not couple to any of the right-handed charged leptons. On the other hand $\bar{\phi}_\alpha \bar{\phi}_\alpha \phi_\beta$, $\bar{\phi}_\alpha \bar{\phi}_\beta \phi_\alpha$ and $\bar{\phi}_\beta \bar{\phi}_\beta \phi_\beta$ which transform as $-i$, couple with $\tau_R$. From the table, it is clear that these products lead to non-vanishing elements in the third position only. The cubic products provide ${\mathcal O}(\epsilon^3)$ contributions to the mass matrix. The aforementioned position corresponds to the position of the ${\mathcal O}(\epsilon^3)$ element in the mass matrix, Eq.~(\ref{eq:leptcontrib}).

\subsection*{Quartic Products}
\begin{flalign}
\text{{\textit {\textbf {i}}})\,\,}\boldsymbol{3}\otimes&\boldsymbol{3}\otimes\boldsymbol{3}\otimes\boldsymbol{3}&\notag\\
&=(\boldsymbol{6}\oplus\boldsymbol{\tb})\otimes(\boldsymbol{6}\oplus\boldsymbol{\tb})&\notag\\
&=\boldsymbol{\xb}\oplus\boldsymbol{\xb}\oplus\boldsymbol{\xb}\oplus\boldsymbol{3}\oplus\boldsymbol{\xb}\oplus\boldsymbol{3^{(0,1)}}\oplus\boldsymbol{3^{(1,0)}}\oplus\boldsymbol{3^{(1,1)}}&\notag\\
&\quad\quad\oplus\boldsymbol{3}\oplus\boldsymbol{\xb}\oplus\boldsymbol{3^{(0,1)}}\oplus\boldsymbol{3^{(1,0)}}\oplus\boldsymbol{3^{(1,1)}}&\notag\\
&\quad\quad\oplus\boldsymbol{3}\oplus\boldsymbol{\xb}\oplus\boldsymbol{3^{(0,1)}}\oplus\boldsymbol{3^{(1,0)}}\oplus\boldsymbol{3^{(1,1)}}&\notag\\
&\quad\quad\oplus\boldsymbol{\xb}\oplus\boldsymbol{3}&\label{eq:quartic1}\\
\text{{\textit {\textbf {ii}}})\,\,}\boldsymbol{3}\otimes&\boldsymbol{3}\otimes\boldsymbol{\tb}\otimes\boldsymbol{\tb}&\notag\\
&=(\boldsymbol{6}\oplus\boldsymbol{\tb})\otimes(\boldsymbol{\xb}\oplus\boldsymbol{3})&\notag\\
&=\boldsymbol{1}\oplus\boldsymbol{8}\oplus\boldsymbol{1^{(0,1)}}\oplus\boldsymbol{1^{(1,0)}}\oplus\boldsymbol{1^{(1,1)}}\oplus\boldsymbol{8}\oplus\boldsymbol{8}&\notag\\
&\quad\quad\oplus\boldsymbol{2}\oplus\boldsymbol{8}\oplus\boldsymbol{8}\oplus\boldsymbol{2}\oplus\boldsymbol{8}\oplus\boldsymbol{8}&\notag\\
&\quad\quad\oplus\boldsymbol{1}\oplus\boldsymbol{8}&\label{eq:quartic2}\\
\text{{\textit {\textbf {iii}}})\,\,}\boldsymbol{\tb}\otimes&\boldsymbol{\tb}\otimes\boldsymbol{3}\otimes\boldsymbol{\tb}&\notag\\
&=(\boldsymbol{\xb}\oplus\boldsymbol{3})\otimes\boldsymbol{3}\otimes\boldsymbol{\tb}&\notag\\
&=(\boldsymbol{\tb}\oplus\boldsymbol{6}\oplus\boldsymbol{\ab}\oplus\boldsymbol{\bb}\oplus\boldsymbol{\cb}\oplus\boldsymbol{6}\oplus\boldsymbol{\tb})\otimes\boldsymbol{\tb}&\notag\\
&=\boldsymbol{\xb}\oplus\boldsymbol{3}\oplus\boldsymbol{3}\oplus\boldsymbol{\xb}\oplus\boldsymbol{3^{(0,1)}}\oplus\boldsymbol{3^{(1,0)}}\oplus\boldsymbol{3^{(1,1)}}&\notag\\
&\quad\quad\oplus\boldsymbol{\xb}\oplus\boldsymbol{3^{(0,1)}}\oplus\boldsymbol{\xb}\oplus\boldsymbol{3^{(1,0)}}\oplus\boldsymbol{\xb}\oplus\boldsymbol{3^{(1,1)}}&\notag\\
&\quad\quad\oplus\boldsymbol{3}\oplus\boldsymbol{\xb}\oplus\boldsymbol{3^{(0,1)}}\oplus\boldsymbol{3^{(1,0)}}\oplus\boldsymbol{3^{(1,1)}}\oplus\boldsymbol{\xb}\oplus\boldsymbol{3}&\label{eq:quartic3}
\end{flalign}
The above expansions, Eqs.~(\ref{eq:quartic1}-\ref{eq:quartic3}), do not contribute to any coupling, since $\boldsymbol{\bar{3}}$ does not appear in their RHS.

\begin{flalign}
\text{{\textit {\textbf {iv}}})\,\,}\boldsymbol{\tb}\otimes&\boldsymbol{\tb}\otimes\boldsymbol{\tb}\otimes\boldsymbol{\tb}&
\end{flalign}
This tensor product corresponds to the conjugation of Eq.~(\ref{eq:quartic1}). The conjugate expansion will have four $\boldsymbol{\tb}$s\footnote{For the sake of brevity, we do not provide the explicit expressions of these quartic $\boldsymbol{\tb}$s. However it is straightforward to obtain them, as was the case for the cubic $\boldsymbol{\tb}$s, Eqs.~(\ref{eq:cubic31}, \ref{eq:cubic32}).} in the RHS. The product $\boldsymbol{\tb}\otimes\boldsymbol{\tb}\otimes\boldsymbol{\tb}\otimes\boldsymbol{\tb}$ can be obtained in terms of the flavons $\phi_\alpha$ and $\phi_\beta$ in several different ways. All these are listed in Table~\ref{tab:quartic1}.  For each combination of flavons, we provide the corresponding $C_4$ representation. Under $\langle\phi_\alpha\rangle\propto(1,0,0)$ and $\langle\phi_\beta\rangle\propto(0,0,1)$, we calculate the vacuum alignments of the above mentioned four $\boldsymbol{\bar{3}}$s and list them in the table.

{\renewcommand{\arraystretch}{1.5}
\begin{table}[]
\begin{center}
\begin{tabular}{|c|c|c c c c|}
\hline
 &$C_4$	&$1^\text{st}\,\boldsymbol{\tb}$	&$2^\text{nd}\,\boldsymbol{\tb}$	&$3^\text{rd}\,\boldsymbol{\tb}$	&$4^\text{th}\,\boldsymbol{\tb}$\\
\hline
$\bar{\phi}_\alpha \bar{\phi}_\alpha \bar{\phi}_\alpha \bar{\phi}_\alpha$
&$1$&$(0,0,0)$&$(0,0,0)$&$(0,0,0)$&$(0,0,0)$\\
$\bar{\phi}_\alpha \bar{\phi}_\alpha \bar{\phi}_\alpha \bar{\phi}_\beta$
&$-1$&$(0,0,\frac{1}{2\sqrt{2}})$&$(0,0,0)$&$(0,0,0)$&$(0,0,0)$\\
$\bar{\phi}_\alpha \bar{\phi}_\alpha \bar{\phi}_\beta \bar{\phi}_\beta$
&$1$&$(0,0,0)$&$(0,0,0)$&$(0,0,0)$&$(0,0,0)$\\
$\bar{\phi}_\alpha \bar{\phi}_\beta \bar{\phi}_\beta \bar{\phi}_\beta$
&$-1$&$(\frac{-1}{2\sqrt{2}},0,0)$&$(0,0,0)$&$(0,0,0)$&$(0,0,0)$\\
$\bar{\phi}_\beta \bar{\phi}_\beta \bar{\phi}_\beta \bar{\phi}_\beta$
&$1$&$(0,0,0)$&$(0,0,0)$&$(0,0,0)$&$(0,0,0)$\\
\hline
\end{tabular}
\end{center}
\caption{Quartic products of $\phi_\alpha$ and $\phi_\beta$ of the form $\boldsymbol{\tb}\otimes\boldsymbol{\tb}\otimes\boldsymbol{\tb}\otimes\boldsymbol{\tb}$ leading to $\boldsymbol{\bar{3}}$s. }
\label{tab:quartic1}
\end{table}
}

\begin{flalign}
\text{{\textit {\textbf {v}}})\,\,}\boldsymbol{3}\otimes&\boldsymbol{3}\otimes\boldsymbol{\tb}\otimes\boldsymbol{3}&
\end{flalign}
This tensor product corresponds to the conjugation of Eq.~(\ref{eq:quartic3}). The conjugate expansion will have four $\boldsymbol{\tb}$s\footnote{We do not provide the expressions of these $\boldsymbol{\tb}$s also.} in the RHS. All the products of $\phi_\alpha$ and $\phi_\beta$ in the form of $\boldsymbol{3}\otimes\boldsymbol{3}\otimes\boldsymbol{\tb}\otimes\boldsymbol{3}$ are listed in Table~\ref{tab:quartic3}, along with their respective $C_4$ representations. Under $\langle\phi_\alpha\rangle\propto(1,0,0)$ and $\langle\phi_\beta\rangle\propto(0,0,1)$, the four $\boldsymbol{\tb}$s attain specific alignments which we have calculated and provided in the table.

{\renewcommand{\arraystretch}{1.5}
\begin{table}[]
\begin{center}
\begin{tabular}{|c|c|c c c c|}
\hline
&$C_4$	&$1^\text{st}\,\boldsymbol{\tb}$	&$2^\text{nd}\,\boldsymbol{\tb}$	&$3^\text{rd}\,\boldsymbol{\tb}$	&$4^\text{th}\,\boldsymbol{\tb}$\\
\hline
$\phi_\alpha \phi_\alpha \bar{\phi}_\alpha \phi_\alpha$
&$-1$&$(0,0,0)$&$(0,0,0)$&$(0,0,0)$&$(0,0,0)$\\
$\phi_\alpha \phi_\alpha \bar{\phi}_\alpha \phi_\beta$
&$1$&$(0,0,0)$&$(0,-\frac{1}{2},0)$&$(0,0,0)$&$(0,0,0)$\\
$\phi_\alpha \phi_\alpha \bar{\phi}_\beta \phi_\alpha$
&$1$&$(0,0,0)$&$(0,0,\frac{-1}{2\sqrt{2}})$&$(0,0,0)$&$(0,0,0)$\\
$\phi_\alpha \phi_\alpha \bar{\phi}_\beta \phi_\beta$
&$-1$&$(0,0,0)$&$(\frac{-1}{2\sqrt{2}},0,0)$&$(0,0,0)$&$(0,0,0)$\\
$\phi_\beta \phi_\beta \bar{\phi}_\alpha \phi_\alpha$
&$-1$&$(0,0,0)$&$(0,0,\frac{1}{2\sqrt{2}})$&$(0,0,0)$&$(0,0,0)$\\
$\phi_\beta \phi_\beta \bar{\phi}_\alpha \phi_\beta$
&$1$&$(0,0,0)$&$(\frac{1}{2\sqrt{2}},0,0)$&$(0,0,0)$&$(0,0,0)$\\
$\phi_\beta \phi_\beta \bar{\phi}_\beta \phi_\alpha$
&$1$&$(0,\frac{1}{2},0)$&$(0,0,0)$&$(0,0,0)$&$(0,0,0)$\\
$\phi_\beta \phi_\beta \bar{\phi}_\beta \phi_\beta$
&$-1$&$(0,0,0)$&$(0,0,0)$&$(0,0,0)$&$(0,0,0)$\\
\hline
\end{tabular}
\end{center}
\caption{Quartic products of $\phi_\alpha$ and $\phi_\beta$ of the form $\boldsymbol{3}\otimes\boldsymbol{3}\otimes\boldsymbol{\tb}\otimes\boldsymbol{3}$ leading to $\boldsymbol{\bar{3}}$s. }
\label{tab:quartic3}
\end{table}
}

The quartic products in Tables~(\ref{tab:quartic1}, \ref{tab:quartic3}) with the $C_4$ representations $1$ and $-1$ couple to $\mu_R$ and $e_R$ respectively. The VEVs with $C_4\equiv 1$ have non-zero elements in the first, the second and the third positions while the VEVs with $C_4\equiv -1$ have non-zero elements in the first and the third positions only. The quartic products provide ${\mathcal O}(\epsilon^4)$ contributions to the mass matrix. The aforementioned positions correspond to the positions of the ${\mathcal O}(\epsilon^4)$ elements in the mass matrix, Eq.~(\ref{eq:leptcontrib}).

%----------------------------------------------------------------------------------------------------------------------------------
\section*{Appendix C: \quad Flavon Potentials}

Here we discuss the flavon potentials that lead to the vacuum alignments assumed in our model. It should be noted that even though our construction results in the required VEVs, we are not doing an exhaustive analysis of the most general flavon potentials involving all the possible invariant terms. However, the content we include is sufficient to realise our VEVs.

\subsection{The triplet flavons: $\phi_\alpha$, $\phi_\beta$}\label{sub:triplet}

First we consider the triplet flavons $\phi_\alpha$ and $\phi_\beta$. Our target is to obtain the VEVs, $\langle\phi_\alpha\rangle=V(1,0,0)^Tm$ and $\langle\phi_\beta\rangle=V(0,0,1)^Tm$, Eqs.~(\ref{eq:leptvev}). The flavons $\phi_\alpha$ and $\phi_\beta$ transform as $\boldsymbol{3}$, Table~\ref{tab:flavourcontent}. The $3\times3$ maximal matrix $V$, Eqs.~(\ref{eq:gen3}), is one of the generators of $\boldsymbol{3}$. Therefore if the potentials of $\phi_\alpha$ and $\phi_\beta$ have minima at $(1,0,0)^Tm$ and $(0,0,1)^Tm$, then they have minima at $V(1,0,0)^Tm$ and $V(0,0,1)^Tm$ as well. The $3\times3$ cyclic matrix $E$, Eqs.~(\ref{eq:gen3}), is another generator of $\boldsymbol{3}$. Therefore, if the potential has a minimum at $(1,0,0)^Tm$, then it has a minimum at $(0,0,1)^Tm$ also. So, for obtaining the target VEVs, all we need to do is to construct a potential with a minimum at $(1,0,0)^Tm$.

An invariant term (singlet) can be constructed using the tensor product expansion of a $\boldsymbol{3}$ and a $\boldsymbol{\tb}$, Eq.~(\ref{eq:tensor33b}). This expansion is valid for both $\Sigma(72\times3)$ and $SU(3)$.  It is well known that the singlet constructed from a triplet and its conjugate is the square of the norm of the triplet, e.g.~for the flavon $\phi_\alpha=(\phi_{\alpha 1},\phi_{\alpha 2},\phi_{\alpha 3})^T$, we have $|\phi_\alpha|^2=\bar{\phi}_{\alpha 1}\phi_{\alpha 1}+\bar{\phi}_{\alpha 2}\phi_{\alpha 2}+\bar{\phi}_{\alpha 3}\phi_{\alpha 3}$. Next we take the symmetric part of the tensor product of two $\boldsymbol{3}$s to obtain a $\boldsymbol{6}$, similar to Eq.~(\ref{eq:Xnu}),
\begin{equation}\label{eq:Salpha}
S_\alpha= \left(\begin{matrix}\phi_{\alpha1}^2\\
	\phi_{\alpha2}^2\\
	\phi_{\alpha3}^2\\
	\sqrt{2}\phi_{\alpha2}\phi_{\alpha3} \\
	\sqrt{2}\phi_{\alpha3}\phi_{\alpha1}\\
\sqrt{2}\phi_{\alpha1}\phi_{\alpha2}
\end{matrix}\right).
\end{equation}
With this sextet, we may construct a singlet by combining it with its conjugate, i.e.~$\bar{S}_\alpha^T S_\alpha$. It can be shown that $\bar{S}_\alpha^T S_\alpha=|\phi_\alpha|^4$. Therefore, without loss of generality we may choose,
\begin{equation}\label{eq:potalpha1}
{\cal T}_{\phi_\alpha}=(|\phi_\alpha|^2-m^2)^2
\end{equation}
as the potential term having a minimum at $(1,0,0)^Tm$. ${\cal T}_{\phi_\alpha}$ is invariant not only under $\Sigma(72\times3)$ but also under the continuous symmetry, $SU(3)$, and hence the minima of the potential are not discrete. This issue can be tackled in two different ways. We may add higher-order non-renormalisable terms which break $SU(3)$ to $\Sigma(72\times3)$. Or we may introduce extra flavons whose sole purpose is to break $SU(3)$ while maintaining renormalisability. In this paper we choose the later approach.

To achieve $SU(3)$ breaking we introduce a flavon $\eta_\alpha=(\eta_{\alpha 1},\eta_{\alpha 2})^T$ which transforms as a doublet under $\Sigma(72\times3)$, Eqs.~(\ref{eq:gen2}) . For $\eta_\alpha$, we construct the following potential:
\begin{equation}\label{eq:potalpha2}
\begin{split}
{\cal T}_{\eta_\alpha}= &(|\eta_\alpha|^2-m^2)^2 \\
&+ \text{Re}^2(\eta_\alpha^Tu_1\,\eta_\alpha) +  \text{Re}^2(\ob\,\eta_\alpha^Tu_{\om}\,\eta_\alpha) + \text{Re}^2(\om\,\eta_\alpha^Tu_{\ob}\,\eta_\alpha),
\end{split}
\end{equation}
where $ \text{Re}^2$ denotes the square of the real part. Since the terms \, $\eta_\alpha^Tu_1\,\eta_\alpha$, \, $\eta_\alpha^Tu_{\om}\,\eta_\alpha$ \, and $\eta_\alpha^Tu_{\ob}\,\eta_\alpha$ \, transform as $\boldsymbol{1^{(0,1)}}$, $\boldsymbol{1^{(1,0)}}$ and $\boldsymbol{1^{(1,1)}}$ respectively, Eqs.~(\ref{eq:tensor22cg1}), their squares are invariants. Therefore it is evident that $ \text{Re}^2(\eta_\alpha^Tu_1\,\eta_\alpha)$, $\text{Re}^2(\ob\,\eta_\alpha^Tu_{\om}\,\eta_\alpha)$ and $\text{Re}^2(\om\,\eta_\alpha^Tu_{\ob}\,\eta_\alpha)$ are also invariants. In terms of the components of $\eta_{\alpha}$, the invariants in Eq.~(\ref{eq:potalpha2}) are given by
\begin{align}
(|\eta_\alpha|^2-m^2)^2&=(\bar{\eta}_{\alpha 1}\eta_{\alpha 1}+\bar{\eta}_{\alpha 2}\eta_{\alpha 2}-m^2)^2,\label{eq:inv21}\\
\text{Re}^2(\eta_\alpha^Tu_1\,\eta_\alpha)&=\text{Re}^2\left(\frac{i\sqrt{2}}{\sqrt{3}}(\eta_{\alpha1}^2-\sqrt{2}\eta_{\alpha1}\eta_{\alpha2}-\eta_{\alpha2}^2)\right),\label{eq:inv22}\\
\text{Re}^2(\ob\,\eta_\alpha^Tu_{\om}\,\eta_\alpha)&=\text{Re}^2\left(\ob\frac{i\sqrt{2}}{\sqrt{3}}(\om \eta_{\alpha1}^2-\sqrt{2}\eta_{\alpha1}\eta_{\alpha2}-\ob\eta_{\alpha2}^2)\right),\label{eq:inv23}\\
\text{Re}^2(\om\,\eta_\alpha^Tu_{\ob}\,\eta_\alpha)&=\text{Re}^2\left(\om\frac{i\sqrt{2}}{\sqrt{3}}(\ob\eta_{\alpha1}^2-\sqrt{2}\eta_{\alpha1}\eta_{\alpha2}-\om\eta_{\alpha2}^2)\right).\label{eq:inv24}
\end{align}
If we assign
\begin{equation}\label{eq:etavev}
\langle\eta_\alpha\rangle = (1,0)^Tm,
\end{equation}
it is evident that each of these invariants vanishes. Therefore the potential, Eq.~(\ref{eq:potalpha2}), attains its minimum value of zero at $\eta_\alpha = (1,0)^Tm$ (and also at the states generated by the discrete transformations on $(1,0)^Tm$). Note that the first term, $(|\eta_\alpha|^2-m^2)^2$, is $SU(2)$ invariant. The other three terms break the continuous $SU(2)$ group and all its $U(1)$ subgroups so that only discrete symmetries generated by Eqs.~(\ref{eq:gen2}) are present in the potential. 

The Kronecker product of $\eta_\alpha (\bold{2})$ and $\bar{\phi}_\alpha (\bold{\tb})$, calculated using Eq.~(\ref{eq:tensor2exp}), leads to a sextet ($\bold{6}$),
\begin{equation}\label{eq:Kalpha}
K_\alpha= \left(\begin{matrix}\eta_{\alpha1} \bar{\phi}_{\alpha1}\\
	\eta_{\alpha1} \bar{\phi}_{\alpha2}\\
	\eta_{\alpha1} \bar{\phi}_{\alpha3}\\
	\eta_{\alpha2} \bar{\phi}_{\alpha1} \\
	\eta_{\alpha2} \bar{\phi}_{\alpha2}\\
\eta_{\alpha2} \bar{\phi}_{\alpha3}
\end{matrix}\right).
\end{equation}
We utilise $S_\alpha$, Eq.~(\ref{eq:Salpha}), and $K_\alpha$, Eq.~(\ref{eq:Kalpha}), to couple together the flavons $\eta_\alpha$, $\phi_\alpha$ and their conjugates and thus we construct
\begin{equation}\label{eq:potalpha3}
{\cal T}_{\phi_\alpha\eta_\alpha}=\left(\bar{S}_\alpha - \bar{K}_\alpha \right)^T\left(S_\alpha - K_\alpha \right)
\end{equation}
as an invariant\footnote{Under the group $C_4$, Table \ref{tab:flavourcontent}, $\phi_\alpha$ belongs to $-i$. Hence $\eta_\alpha$ needs to transform as $i$ to ensure the invariance of Eq.~(\ref{eq:potalpha3})}. If we assign 
\begin{equation}\label{eq:phivev}
\langle\phi_\alpha\rangle = (1,0,0)^Tm,
\end{equation}
its symmetric product, $S_\alpha$,  becomes $(1,0,0,0,0,0)^Tm^2$. The Kronecker product, $K_\alpha$, of $\langle\eta\rangle=(1,0)^Tm$ and $\langle\phi_\alpha\rangle=(1,0,0)^Tm$ also becomes $(1,0,0,0,0,0)^Tm^2$. Therefore, ${\cal T}_{\phi_\alpha\eta_\alpha}$ vanishes (which is its minimum value) at assigned VEVs, Eqs.~(\ref{eq:etavev}, \ref{eq:phivev}).

Combining Eqs.~(\ref{eq:potalpha1}, \ref{eq:potalpha2}, \ref{eq:potalpha3}), we obtain the following renormalisable potential term for the flavon $\phi_\alpha$:
\begin{equation}\label{eq:potetaalpha}
{\cal T}_{\phi_\alpha}+{\cal T}_{\eta_\alpha}+{\cal T}_{\phi_\alpha\eta_\alpha}
\end{equation}
which is $\Sigma(72\times3)$ invariant and at the same time devoid of continuous symmetries. A similar potential can be constructed for the flavon $\phi_\beta$ also,
\begin{equation}\label{eq:potetabeta}
{\cal T}_{\phi_\beta}+{\cal T}_{\eta_\beta}+{\cal T}_{\phi_\beta\eta_\beta},
\end{equation}
by introducing a doublet $\eta_\beta$\footnote{Under the group $C_4$, Table \ref{tab:flavourcontent}, $\phi_\beta$ belongs to $i$. Hence $\eta_\beta$ needs to transform as $-i$ to ensure the invariance of ${\cal T}_{\phi_\beta\eta_\beta}$ in Eq.~(\ref{eq:potetabeta})}. The expressions for the three invariants in Eq.~(\ref{eq:potetabeta}) can be found by replacing $\alpha$ with $\beta$ in Eqs.~(\ref{eq:Salpha}-\ref{eq:inv24}, \ref{eq:Kalpha}, \ref{eq:potalpha3}). We also write the term,\begin{equation}
{\cal T}_{\phi_\alpha\phi_\beta}=|\phi_\alpha^\dagger \phi_\beta|^2,
\end{equation}
which couples $\phi_\alpha$ and $\phi_\beta$ together and ensures that their VEVs are orthogonal to each other, Eqs.~(\ref{eq:leptvev}). In conclusion, the potential,
\begin{equation}\label{eq:alphabeta}
{\cal T}_{\phi_\alpha}+{\cal T}_{\phi_\beta}+{\cal T}_{\eta_\alpha}+{\cal T}_{\eta_\beta}+{\cal T}_{\phi_\alpha\eta_\alpha}+{\cal T}_{\phi_\beta\eta_\beta}+{\cal T}_{\phi_\alpha\phi_\beta}
\end{equation}
which is invariant under $\Sigma(72\times3)$, has a discrete set of minima. One among them corresponds to the required VEVs, Eqs.~(\ref{eq:leptvev}). The flavons $\phi_\alpha$ and $\phi_\beta$ attain these VEVs through the spontaneous symmetry breaking of $\Sigma(72\times3)$.

\subsection{The sextet flavon: $\xi$}\label{sub:sextet}

We studied the invariants that can be constructed using the sextet $\xi$ up to the quartic order (renormalisable) and found that these terms are insufficient to obtain a potential devoid of continuous symmetries ($SU(3)$ and its continuous subgroups). Therefore, as in Subsection~\ref{sub:triplet}, we introduce extra flavons to break the continuous symmetries and to ensure that the potential has a discrete set of minima. The extra flavons introduced here are a doublet $\eta$ and two triplets $\phi_a$, $\phi_b$. The flavons used in the charged-lepton sector ( $\phi_\alpha$, $\phi_\beta$, $\eta_\alpha$, $\eta_\beta$ in Subsection~\ref{sub:triplet}) are kept distinct from the flavons used in the neutrino sector ($\xi$, $\eta$, $\phi_a$, $\phi_b$ in Subsection~\ref{sub:sextet}) in order to avoid unwanted couplings between the two sectors. Table~\ref{tab:fullcontent} provides the complete list of flavons in the model along with the fermions. 
{\renewcommand{\arraystretch}{1.5}
\begin{table}[]
\begin{center}
\begin{tabular}{|c|c c c c c c c c|}
\hline
 	&$e_R$	&$\mu_R$&$\tau_R$&$L$	&$\nu_R$&$\phi_\alpha$&$\phi_\beta$&$\xi$\\
\hline
$\Sigma(72\times3)$&$\boldsymbol{1}$&$\boldsymbol{1}$&$\boldsymbol{1}$&$\boldsymbol{\tb}$&$\boldsymbol{\tb}$&$\boldsymbol{3}$&$\boldsymbol{3}$&$\boldsymbol{6}$\\
$C_4$	&$-1$&$1$&$i$&$1$&$1$&$-i$&$i$&$1$\\
$C_3$	&$\om$&$\om$&$\om$&$\om$&$\om$&$1$&$1$&$\om$\\
\hline
	&$\eta_\alpha$&$\eta_\beta$&$\eta$&$\phi_a$&$\phi_b$& & &\\
\hline
$\Sigma(72\times3)$&$\boldsymbol{2}$&$\boldsymbol{2}$&$\boldsymbol{2}$&$\boldsymbol{3}$&$\boldsymbol{3}$& & &\\
$C_4$&$i$&$-i$&$1$&$1$&$1$& & &\\	
$C_3$&$1$&$1$&$1$&$\ob$&$\ob$& & &\\
\hline
\end{tabular}
\end{center}
\caption{The full flavour structure of the model. The flavons in the upper half, $\phi_\alpha$, $\phi_\beta$, $\xi$, are the ones whose VEVs form the charged-lepton and neutrino mass matrices. The lower half comprises extra flavons added to break the continuous symmetries in the potentials. The sole purpose of $C_3$ is to avoid unwanted couplings between the charged-lepton and the neutrino sectors.}
\label{tab:fullcontent}
\end{table}
}

Our first step is to write the potential terms for $\eta$, $\phi_a$ and $\phi_b$, similar to Eq.~(\ref{eq:alphabeta}),
\begin{equation}\label{eq:ab}
{\cal T}_{\phi_a}+{\cal T}_{\phi_b}+{\cal T}_{\eta}+{\cal T}_{\phi_a\eta}+{\cal T}_{\phi_b\eta}+{\cal T}_{\phi_a\phi_b}.
\end{equation}
The individual invariant terms in Eq.~(\ref{eq:ab}) are
\begin{flalign}
{\cal T}_{\phi_a}&=(|\phi_a|^2-m^2)^2,\label{eq:abterms1}\\
{\cal T}_{\phi_b}&=(|\phi_b|^2-m^2)^2,\\
{\cal T}_{\eta}&=(|\eta|^2-m^2)^2 \\
&\quad+ \text{Re}^2(\eta^Tu_1\,\eta) +  \text{Re}^2(\ob\,\eta^Tu_{\om}\,\eta) + \text{Re}^2(\om\,\eta^Tu_{\ob}\,\eta),\notag\\
{\cal T}_{\phi_a\eta}&=\left(\bar{S}_a - \bar{K}_a \right)^T\left(S_a - K_a \right),\\
{\cal T}_{\phi_b\eta}&=\left(\bar{S}_b - \bar{K}_b \right)^T\left(S_b - K_b \right),\\
{\cal T}_{\phi_a\phi_b}&=|\phi_a^\dagger \phi_b|^2\label{eq:abterms7},
\end{flalign}
where $S_a$, $K_a$ and $S_b$, $K_b$ are defined similar to $S_\alpha$, $K_\alpha$ in Eqs.~(\ref{eq:Salpha}, \ref{eq:Kalpha}) having $\phi_\alpha$, $\eta_\alpha$ replaced with $\phi_a$, $\eta$ and $\phi_b$, $\eta$ respectively. As described earlier, it is straightforward to show that, each term in Eqs.~(\ref{eq:abterms1}-\ref{eq:abterms7}) vanishes, if we assign the following VEVs:
\begin{align}
\langle\eta\rangle =& (1,0)^Tm,\\
\langle\phi_a\rangle = &(1,0,0)^Tm,\label{eq:abvevsa}\\
\langle\phi_b\rangle = &(0,0,1)^Tm.\label{eq:abvevsb}
\end{align}

$S_a$ and $S_b$ are the sextets constructed from $\phi_a$ and $\phi_b$ respectively. We may also construct a sextet combining $\phi_a$ and $\phi_b$ together, 
\begin{equation}\label{eq:Sab}
S_{ab}= \left(\begin{matrix}\phi_{a1} \phi_{b1} \\
	\phi_{a2} \phi_{b2}\\
	\phi_{a3} \phi_{b3}\\
	\frac{1}{\sqrt{2}}\left(\phi_{a2} \phi_{b3} + \phi_{a3} \phi_{b2}\right)\\
	\frac{1}{\sqrt{2}}\left(\phi_{a3} \phi_{b1} + \phi_{a1} \phi_{b3}\right)\\
\frac{1}{\sqrt{2}}\left(\phi_{a1} \phi_{b2} + \phi_{a2} \phi_{b1}\right)
\end{matrix}\right).
\end{equation}
Under the VEVs, Eqs.~(\ref{eq:abvevsa}, \ref{eq:abvevsb}), we obtain
\begin{flalign}
\langle S_a\rangle =& (1,0,0,0,0,0)^Tm^2,\label{eq:savev}\\
\langle S_b\rangle =& (0,0,1,0,0,0)^Tm^2,\label{eq:sbvev}\\
\langle S_{ab}\rangle = &(0,0,0,0,\frac{1}{\sqrt{2}},0)^Tm^2.\label{eq:sabvev}
\end{flalign}
%xxxxxxxxxxxxxxxxxxxxxxxxxxxxxxxxxxxxxxxxxxxxxxxxx

We use the sextet $\xi=(\xi_1 ,\xi_2 ,\xi_3 ,\xi_4 ,\xi_5, \xi_6)$ to construct the Majorana neutrino mass matrix, Eq.~(\ref{eq:Tnu}). In the VEV of $\xi$, if any two among the three elements $\xi_4$, $\xi_5$ and $\xi_6$ become zero, then two off-diagonal elements in the mass matrix vanishes. $U_\nu$ effectively becomes a $2\times2$ unitary matrix and $U_{\text PMNS}=VU_\nu$ attains one trimaximal column. In all the four VEVs, Eqs.~(\ref{eq:vevtxmp}, \ref{eq:vevtxmm}, \ref{eq:vevtpmp}, \ref{eq:vevtpmm}), we can see that $\xi_4=0$ and $\xi_6=0$ leading to the trimaximal second column, i.e.~$|U_{e2}|=|U_{\mu2}|=|U_{\tau2}|=\frac{1}{\sqrt{3}}$. Both $\txm$ and $\tpm$ belong to the larger class of mixing schemes in which one neutrino is trimaximally mixed\cite{Harrison03}. As the first step in constructing the potential for $\xi$, we consider the tensor product of two $\xi$s, Eq.~(\ref{eq:tensor66}), and obtain a sextet,
\begin{equation}\label{eq:Xxi}
\bar{X}=\left(\begin{matrix}\frac{1}{\sqrt{3}}\xi_{\{2}\xi_{3\}}-\frac{1}{\sqrt{3}}\xi_4 \xi_4\\
\frac{1}{\sqrt{3}}\xi_{\{3}\xi_{1\}}-\frac{1}{\sqrt{3}}\xi_5 \xi_5\\
\frac{1}{\sqrt{3}}\xi_{\{1}\xi_{2\}}-\frac{1}{\sqrt{3}}\xi_6 b_6\\
\frac{1}{\sqrt{6}}\xi_{\{5}\xi_{6\}}-\frac{1}{\sqrt{3}}\xi_{\{1}\xi_{4\}}\\
\frac{1}{\sqrt{6}}\xi_{\{6}\xi_{4\}}-\frac{1}{\sqrt{3}}\xi_{\{2}\xi_{5\}}\\
\frac{1}{\sqrt{6}}\xi_{\{4}\xi_{5\}}-\frac{1}{\sqrt{3}}\xi_{\{3}\xi_{6\}}\\
\end{matrix}\right), 
\end{equation}
as shown in Eq.~(\ref{eq:tensor66exp1}). $\bar{X}$ transforms as a $\boldsymbol{\xb}$. Note that, when $\xi_4=0$ and $\xi_6=0$, the fourth and sixth elements of $\bar{X}$ also vanishes.

If the second column of $U_{\text PMNS}$ is trimaximally mixed, then the VEV, $\langle\xi\rangle$, as well as the resulting $\langle\bar{X}\rangle$ have non-zero elements only in the first, second, third and the fifth positions. As shown in Eqs.~(\ref{eq:savev}, \ref{eq:sbvev}, \ref{eq:sabvev}), $\langle S_a\rangle$, $\langle S_b\rangle$ and $\langle S_{ab}\rangle$ have non-zero elements only in the first, third and the fifth position respectively. Therefore, a linear combination of $\langle\xi\rangle$, $\langle X_{\xi}\rangle$, $\langle S_a\rangle$, $\langle S_b\rangle$ and $\langle S_{ab}\rangle$ can be constructed which fully vanishes. With this information in hand, we construct the potential term,
\begin{equation}\label{eq:potxi}
\begin{split}
{\cal T}_{\xi}= &\left(m\,\bar{\xi} + \bar{c}_1 \bar{X} + \bar{c}_2\bar{S}_a + \bar{c}_3\bar{S}_b + \bar{c}_4\bar{S}_{ab} \right)^T \\
&\quad\quad\quad\quad\left(m\,\xi + c_1 X + c_2 S_a + c_3 S_b + c_4 S_{ab} \right),
\end{split}
\end{equation}
where $c_1$, $c_2$, $c_3$ and $c_4$ are constants. ${\cal T}_{\xi}$ couples the sextet flavon, $\xi$ with the triplet flavons, $\phi_a$ and $\phi_b$. Any neutrino mass matrix which leads to a trimaximally-mixed column can be obtained using a potential of the form, Eq.~(\ref{eq:potxi}). The values of the constants resulting in the four VEVs, Eqs.~(\ref{eq:vevtxmp}, \ref{eq:vevtxmm}, \ref{eq:vevtpmp}, \ref{eq:vevtpmm}), are given in Table~\ref{tab:constants}. 

Using an appropriate choice of the constants, $c_1$, $c_2$, $c_3$, and $c_4$, we may obtain any mixing scheme within the constraint of a trimaximal column. It can be shown that, having the symmetry of $c_2$ and $c_3$ being real (invariant under complex conjugation) leads to $\txm$. In the case of $\tpm$, the symmetry is the simultaneous conjugation and interchange of $c_2$ and $c_3$. Additionally, the fact that $c_1$, $c_2$, $c_3$ and $c_4$ are related by simple ratios points to the presence of more symmetries, the study of which is beyond the scope of this paper. 

{\renewcommand{\arraystretch}{1.5}
\begin{table}[]
\begin{center}
\begin{tabular}{|c|c c c c|}
\hline
	&$c_1$	&$c_2$	&$c_3$	&$c_4$\\
\hline
Eq.~(\ref{eq:vevtxmp})		&$\sqrt{3}$		&$-\sqrt{2}t$		&$-2\sqrt{2}t$		&$\sqrt{2}$\\
Eq.~(\ref{eq:vevtxmm})		&$\sqrt{3}$		&$-2\sqrt{2}t$		&$-\sqrt{2}t$		&$\sqrt{2}$\\
Eq.~(\ref{eq:vevtpmp})		&$-\sqrt{3}$		&$\frac{1}{\sqrt{2}}(1-i 3t)$		&$\frac{1}{\sqrt{2}}(1+i 3t)$		&$-3\sqrt{2}t$\\
Eq.~(\ref{eq:vevtpmm})		&$-\sqrt{3}$		&$\frac{1}{\sqrt{2}}(1+i 3t)$		&$\frac{1}{\sqrt{2}}(1-i 3t)$		&$-3\sqrt{2}t$\\
\hline
\end{tabular}
\end{center}
\caption{The values of constants appearing in the potential, Eq.~(\ref{eq:potxi}), for the sextet flavon, $\xi$, corresponding to the four cases. We have  $t=\text{tan}(\frac{\pi}{8})=\sqrt{2}-1$.}
\label{tab:constants}
\end{table}
}

With the help of first and second order partial derivatives of a given potential, its minima can be calculated, as was followed in previous works, e.g.~ in Ref.~\cite{King10}. Using such a procedure, along with numerical analysis, we have verified that every potential discussed here has a discrete set of minima and that the quoted VEVs are included among those minima in each case. 

\providecommand{\href}[2]{#2}\begingroup\raggedright\endgroup

\end{document}